\documentclass[intlimits,twoside,a4paper]{article}

\usepackage{amsmath,amssymb}
\usepackage{graphicx}

\usepackage[T2A]{fontenc}
\usepackage[cp1251]{inputenc}

\usepackage[eqsecnum]{cmpj2}
\usepackage{slantsc}
\usepackage{textcomp}

\newcommand{\be}{\begin{equation}}
\newcommand{\ee}{\end{equation}}
\newcommand{\bea}{\begin{eqnarray}}
\newcommand{\eea}{\end{eqnarray}}

\issue{2014}{17}{4}{43001}
\doinumber{10.5488/CMP.17.43001}

\title[Phase transition of the first order]%
{The phase transition of the first order in the critical region of the gas--liquid system}

  \author{I.R. Yukhnovskii}
  \address{Institute for Condensed Matter Physics of the National Academy
  of Sciences of Ukraine, \\ 1 Svientsitskii St., 79011 Lviv, Ukraine}

\date{Received July 15, 2014, in final form October 1, 2014}
\authorcopyright{I.R. Yukhnovskii, 2014}

\begin{document}

\maketitle

\begin{abstract}
This is a summarising investigation of the events of the phase transition of the first order that occur in the critical region below the liquid--gas critical point.
The grand partition function has been completely integrated in the phase-space of the collective variables. The basic density measure is the quartic one. It has the form of the exponent function with the first, second, third and fourth degree of the collective variables. The problem has been reduced to the Ising model in an external field, the role of which is played by the generalised chemical potential $\mu^*$.
The line $\mu^*(\eta) =0$, where $\eta$ is the density, is the line of the phase transition. We consider the isothermal compression of the gas till the point where the phase transition on the line $\mu^*(\eta) =0$ is reached. When the path of the pressing reaches the line $\mu^* =0$ in the gas medium, a droplet of liquid springs up. The work for its formation is obtained, the surface-tension energy is calculated. On the line $\mu^* =0$ we have a two-phase system: the gas and the liquid (the droplet) one. The equality of the gas and of the liquid chemical potentials is proved.
The process of pressing is going on. But the pressure inside the system has stopped, two fixed densities have arisen: one for the gas-phase $\eta_\textrm{g} = \eta_\textrm{c} ( 1 - {d}/{2})$ and the other for the liquid-phase $\eta_\textrm{l} = \eta_\textrm{c} (1 + {d}/{2} )$ (symmetrically to the rectlinear diameter), where $\eta_\textrm{c} = 0.13044$ is the critical density. Starting from that moment the external pressure works as a latent work of pressure. Its value is obtained. As a result, the gas-phase disappears and the whole system turns into liquid. The jump of the density is equal to $\eta_\textrm{c} d$, where $d = \sqrt{{D}/{2G}} \sim \tau^{\nu/2}$.  $D$ and $G$ are coefficients of the Hamiltonian in the last cell connected with the renormalisation-group symmetry. The equation of state is written.
\pacs 05.70.Jk, 64.70.F-, 64.60.F-
\end{abstract}

\section{Introduction}
\label{sec:intro}
The present research follows our previous consideration concerning the liquid--gas critical point. In this work we continue the investigations of the system of interacting particles in the vicinity of the critical point~\cite{Yukh_Kol_Idz_13,Yukh_Idz_Kol_95,Yukh_87,Yukh_nuovo89}. The great partition function was calculated. The potential of the interaction between particles consists of the sum of the van der Waals interaction and of the short-range interaction between the hard-spheres.
We used the collective variables (CV) method.  The Jacobian of the transition from the Cartesian-coordinates space to the space of the collective variables was averaged with the statistical weight of a distribution function of the hard-spheres.

The problem was reduced to the Ising model in an external field, presented by the generalised chemical potential $\mu^*$.

As a basic density measure, the quartic distribution function is used.
It has the form of the exponent function with the first, second, third and
fourth degree of the collective variables. Detailed calculations were
performed in~\cite{Yukh_Idz_Kol_95} for the case of $\tau \geqslant 0$, where
$\tau = (T-T_\textrm{c})/T_\textrm{c}$. Therein, the coordinates of the critical point $T_\textrm{c}$
and $\eta_\textrm{c}$ were obtained, where $T_\textrm{c}$ is the critical temperature, $\eta_\textrm{c}$
is the critical density $\eta_\textrm{c} = \left( {N}/{V} \right)_\textrm{c} ({\pi \sigma^3}/{6})$,
$\sigma$ is the diameter of a hard sphere.

The critical indexes correspond to that of the Ising model. Their values,
the sizes of the particles, the critical temperatures and densities are given
in tables~\ref{table1} and~\ref{tab2} \cite{Yukh_Kol_Idz_13}.

\begin{table}[!t]
	\caption{The critical indexes for the quartic and sextic density measures in the CV method~\cite{Mic_Theory_01}.}
	\begin{center}
		\begin{tabular}{|c|c|c|c|c|}
			\hline\hline
			Critical parameters & $\nu$ & $\alpha$ & $\beta$ & $\gamma$ \\
			\hline\hline
			Approximation $\rho^4$ & 0.605 & 0.185 & 0.303 & 1.210 \\
			Approximation $\rho^6$ & 0.637 & 0.088 & 0.319 & 1.275 \\
			\hline\hline
		\end{tabular}
	\end{center}
	\label{table1}
\end{table}

\begin{table}[!b]	
	\caption{The critical temperatures $T_\textrm{c}$ and the effective hard sphere diameter $\sigma$ for some systems.}
	\begin{center}
		\begin{tabular}{|c|c|c|c|c|c|c|c|c|}
			\hline\hline
			System & $P_\textrm{c}$,  atm& $P_\textrm{c}$,  atm& $T_\textrm{c}$, \textcelsius & $T_\textrm{c}$, \textcelsius &
$\sigma_0$, {\AA} & $\sigma$, {\AA} & $\sigma/\sigma_0$ & $\varepsilon/k_\textrm{B}$, K \\
			&    & ~\cite{Yukh_Kol_Idz_13} &  (exp.) & ~\cite{Yukh_Kol_Idz_13} & (exp.) &   &  &   \\
			\hline\hline
			CO-CO       &34.532&37.92 & --140.23 & --138.46 & 3.76 & 3.37 & 0.898 & 100.2  \\
			Ar-Ar       &47.964&51.75 & --122.65 & --123.27 & 3.405& 3.14 & 0.922 & 119.8  \\
			Kr-Kr       &54.182&56.76 & --63.1   & --67.84  & 3.6  & 3.367 &0.935 & 171    \\
			Xe-Xe       &57.537&65.88 & 16.62   & 16.84   & 4.1  & 3.71 & 0.905 & 221    \\
			O$_2$-O$_2$ &49.77&54.76 & --118.84 & --110.8  & 3.58 & 3.18 & 0.89  & 117.5  \\
			N$_2$-N$_2$ &33.55&44.08 & --147.05 & --150.02 & 3.698& 3.365 &0.91  & 95.05  \\
			\hline\hline
		\end{tabular}
	\end{center}
	\label{tab2}
\end{table}

The first results of the investigations in the area of temperatures below
the critical point, $\tau \leqslant 0$, are given in~\cite{Yukh_Kol_Idz_13,Yukh_Idz_Kol_95,Yukh_87,Yukh_nuovo89}.
It has been shown that the events connected with the phase transition of the first order from the
gas phase to the liquid phase (and inversely) are described by the integral over the variable
$\rho_0$ for ${\bf k} =0$. This integral embraces the results of the integration over all
collective variables with the exception of the variable $\rho_0$.

The collective variable $\rho_0$ is connected with the distribution of the number of particles.
The integration over the variable $\rho_0$ was performed in the steepest descent method.
It is shown that the discriminant $Q$ of the equation concerning the maximum of integrand
can take  positive or negative values, or be equal to zero. When the  discriminant $Q$ is
positive we have pure monophasic states of the gas or of the liquid. In the area $Q < 0$,
the maximization problem possesses three real roots,
\[
\rho_0^{\rm max} = b \cos \frac{\varphi + 2\pi n}{3}\,, \qquad
n = 0, 1, 2, \qquad 0 \leqslant \varphi \leqslant \pi .
\]
However, only two of them, actually for  $n=0$ and $n=1$, correspond to a maximum of the integrand.
The areas of the principal and of the secondary maxima were determined. It was shown that there
is a break between the end of the principal maximum for the gas phase and the beginning of the
principal maximum for the liquid phase.
Those were the main results of the papers~\cite{Yukh_Kol_Idz_13,Yukh_Idz_Kol_95}.

In this work we would like to pass the whole route of the phase transition of the first order.
The work consists of the introduction and seven sections. In Introduction we define the problem
and give the main results of the mentioned works. In section~\ref{sec:monostates} the description
of the monophasic states is given, which corresponds to the positive values of the discriminant $Q$.
Section~\ref{sec:ref_system} is devoted to the coordinate form of the reference system. We work within
the framework of the quartic density measure. The reference system is included in our problem through
the cumulants $\mathfrak{M}_1$, $\mathfrak{M}_2$, $\mathfrak{M}_3$ and $\mathfrak{M}_4$. In accordance with the accuracy
of the consideration of the whole problem, it is needed to take a suitable expression for the pressure $P_0$,
the chemical potential $\mu_0$ and the free energy $F_0$ of the reference system.

Section~\ref{sec:phase_trans} is an essential one in this work. The domain of the phase transition is
considered. Here, the discriminant is negative $Q < 0$. In this region we have two maxima~---
the principal and the secondary one. The principal maxima correspond to the pure gas or to
the pure liquid phases. The secondary maxima are connected with the stable density fluctuations,
which are either of the liquid type in the gas phase, or of the gas type in the liquid phase.

Under consideration is the equation for the envelope curve of two maxima, the principal and the secondary ones:
\[
\frac{PV}{\Theta} - \ln \Xi_{\rho_0} = 0, \qquad
\frac{\partial \ln \Xi_{\rho_0}}{\partial \beta \mu} = N.
\]
Having opened the second equation we get the probabilities of the system to be in the
states corresponding to the principal or to the secondary maxima. We have got analytical
expressions for the probabilities to be in the gas or in the liquid phases.

The central considerations concern the events along the line $\mu^* = 0$. Here,
when the gas system is pressed, a droplet of liquid appears inside it, that is, a
new phase arises~--- the phase of liquid. In the appearance of the droplet lies the
essence of the phase transition of the first order (as a result of isothermal
pressure on the gas system). The work for the formation of a droplet~--- the surface-tension
energy~--- was calculated.

A fundamental quantity $\Delta$ is introduced. It describes an inverse to the usual
order parameter, characterizing the weight of the pure monophase state in the system.
On the rectangle vertex of the $Q = 0$, the weight is equal to unity for the gas phase
and zero for the liquid phase or, on the contrary, dependent on which of them the maternal is.
The quantity $\Delta$ is equal to zero on the line of the phase transition. This  means that
on that line the gas phase (maternal) and the liquid phase (in a droplet-form) have equal
statistical weights. In its initial determination $\Delta$ was a function of the reference
system cumulants $\mathfrak{M}_2$, $\mathfrak{M}_3$,  $\mathfrak{M}_4$ and, therefore, was a function of
the compressibility coefficients and its derivatives. The equality $\Delta =0$ means that
on the line  $\mu^* = 0$, the compressibility coefficient equals zero. We have some
metastable state~--- the state of phase transition of the first order.

In section~\ref{sec:chem_equiv}, the equality of chemical potential of the gas phase and
of the liquid phase on the line $\mu^* = 0$ is shown.

In section~\ref{sec:condens_proc}, the gas-condensation process is analyzed. An analytical
expression for the surface-ten\-sion energy is obtained. The latent work $A_{\textrm{lat}}$ in
the isothermal pressure process (an analogue to the latent heat in an isobaric process),
which acts along the line
$\mu^* = 0$ during the gas-liquid phase transition, was obtained.

And, in the end, in section~\ref{sec:state_equation}, the equation of state was obtained.
The equation for the isotherm $P = P(\eta)_{\tau = {\rm const}}$ has three parts:
\begin{enumerate}

  \item a monophasic gas branch: $P = P(\eta, \tau)$ for densities
$\eta \leqslant \eta_\textrm{g} = \eta_\textrm{c} (1 - d/2)$, $d = \sqrt{{D}/2{G}}$ ($D$ and $G$ are coefficients in Hamiltonian at the second and at the
fourth degree of $\rho_0$);

  \item a horizontal line $\mu^*(\eta) = 0$ for the two-phase system,
here $P = 0$, and $\eta_\textrm{g} \leqslant \eta \leqslant \eta_\textrm{l}$, $\eta_\textrm{l} = \eta_\textrm{c} (1 + d/2)$.
The transition of gas into liquid is takes place at the expense of the latent work $A_{\textrm{lat.}}$ acting from without;

  \item the liquid-branch for $\eta > \eta_\textrm{l} = \eta_\textrm{c} (1 + d/2)$.
\end{enumerate}

Thus, during the gas-condensation process we get a jump of the density of the amount of
$\eta_\textrm{l} - \eta_\textrm{g} = \eta_\textrm{c} d \sim \tau^{\nu/2}$. As we see, the magnitudes of the densities
of the beginning and of the end of the transition process from the gas state to the liquid
state are located symmetrically to the rectilinear diameter.

When the system approaches to the critical point, the specific surface-tension energy
disappears as $\tau^{5/2\nu}$, because $D \sim \tau^{2\nu}$, $G \sim \tau^\nu$, and the jump
of the density $\eta$ is proportional to  $\tau^{\nu/2}$.

In the preamble we stressed that within the framework of the Gibbsian statistics one can describe
the phase transition of the first order~\cite{Martynov:1999en}.

As for other works, we benefited from the monographs of R.~Balescu \cite{balescu1975equilibrium},
L.D.~Landau and E.M.~Lifshits \cite{Land_05}, J.-P.~Hansen  and I.R.~McDonald \cite{Hansen06}
as well as from~\cite{Derz_Myg_e,Pab_Yan_Esc_99,Anisimov_CMP_13}.

In this research we made a number of assumptions, especially concerning the reference system,
the dependence of the coefficients $D$ and $G$ on the density and others. In fact, by means
of the method of collective variables it is possible to describe the critical point theory
of the liquid-gas system with a more sufficient accuracy. Now we get to the theme.

\subsection{How it should look like from the very start}

Let a big volume $V_0$ in the form of a sphere with the radius $R_0$, $V_0 = \frac{4\pi}{3}R_0^3$
and with the number of particles $N$, $N = n_0V_0$ be picked out in a gas medium with the density $n_0$,
pressure $P$ at the temperature~$T$.

We consider the phase transition of the first order as a result of the quasistatic pressure on the gas sphere.
The process takes place at temperatures close to the critical point $T_\textrm{c}$. The dimensionless temperature $\tau$
is introduced $\tau = (T - T_\textrm{c})/{T_\textrm{c}}$, and $|\tau| \leqslant 0,01$, $\tau < 0$. The pressure inside the sphere
of radius $R_0$ is equal to $P + \delta P$, where $\delta P$ is the additional pressure, connected
with the spherical form of the volume $V$. The gas energy inside the sphere:
\be
\label{eq1}
{\cal E}_0 = (P + \delta P)V_0 - \alpha S_0\,,
\ee
where $S_0$ is the area of the sphere, $S_0 = 4\pi R_0^2$, $\alpha$ is the surface-tension coefficient.
The minimum of the difference ${\cal E}_0 - P V_0$ means that
${\delta({\cal E}_0 -PV_0)}/{\delta R} = 0$ or
\be
\label{eq2}
\delta P_0 = \frac{2 \alpha}{R_0}\,.
\ee
Such a picture may be useful to remember when discussing the isothermal phase transition in the
gas-liquid system illustrated in this work.

We consider the grand partition function
\be
\label{eq3}
\Xi = \sum\limits_{N=0}^{\infty} \frac{z^N}{N!}Z_N,
\ee
where $N$ is the number of particles, $z^N$ is the activity
\be
\label{eq4}
z^N = \left[ \left( \frac{mk_\textrm{B}T}{2\pi}\right)^{3/2} \frac{1}{\hbar^3} \right]^N \exp (\beta \mu N),
\ee
$m$ is the mass of a particle, $k_\textrm{B}$ is Bolzman's constant, $T$ is temperature, $\hbar$ is Plank's
constant, $\beta = (k_\textrm{B}T)^{-1}$, $\mu$ is the chemical potential, $Z_N$ is the configuration integral
of  $N$ particles~\cite{Yukh_Kol_Idz_13,Yukh_Idz_Kol_95}.  The
collective-variables method is used $\{\rho_{\bf k}\}$~\cite{Yukh_87,Yukh_nuovo89,Mic_Theory_01} to calculate  $\Xi$.
The Jacobian of transition from the Cartesian coordinates to the collective variables
was calculated with the statistical weight of the system of hard cores.
The latter was taken as a reference system. As a result, for $\Xi$ we get the expression \cite{Yukh_Kol_Idz_13,Yukh_Idz_Kol_95}:
\be
\label{eq5}
\Xi = \Xi_0\Xi_1\Xi_{p_0}\,,
\ee
where $\Xi_0$ is the partition function of the reference system (RS),
\be
\label{eq6}
\ln \Xi_0 = \beta \mu_0 \langle N \rangle - \beta F_0 = \frac{p_0 V_0}{\Theta}\,,
\ee
$\mu_0$ is the chemical potential of RS, $F_0$ is free energy of RS.

\begin{figure}[!t]
	\centering
	\includegraphics[width=0.6\linewidth]{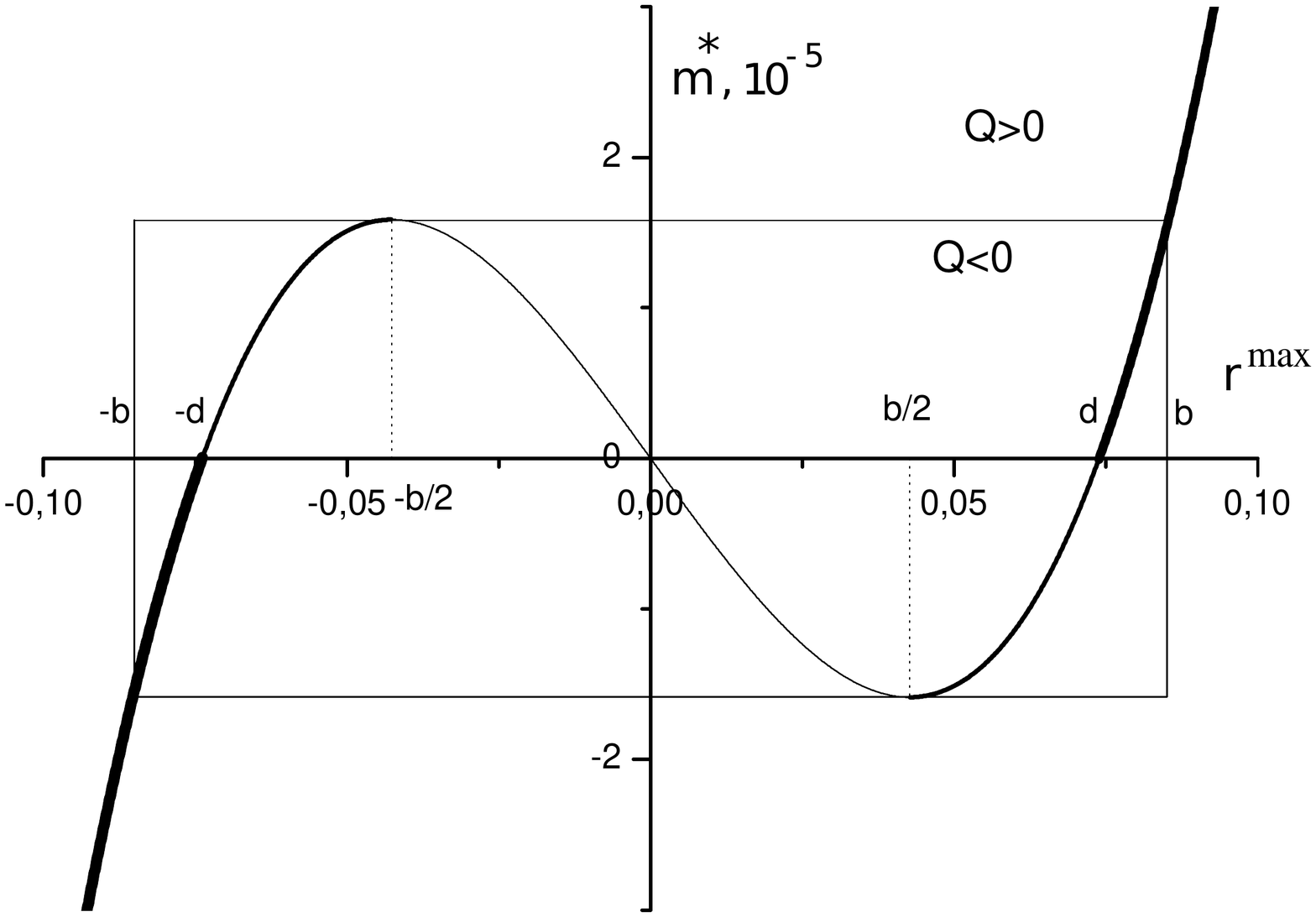}
	\caption{The curve of the generalized chemical potential $\mu^*$ as
functions of $\rho_0^{\rm max}$. On this curve, the integrand $\exp[NE(\rho)]$
in the region of~(\ref{eq8}) has maxima [solutions~(\ref{eq26}) for $Q>0$ and~(\ref{eq27}) for $Q<0$]. Here,
		$a = G \big(\sqrt{{2D}/{3G}}\big)^3 \sim \tau^{5/2\nu}$; $b = \sqrt{{2D}/{3G}}
\sim \tau^{\nu/2}$; $d = \sqrt{ {D}/{2G} } \sim \tau^{\nu/2}$. The thick lines
correspond to principal maxima of the function $E(\mu^*, \rho_0)$, thin lines correspond to
secondary maxima, and the very thin line is a line of minimum. The origin of the system's
coordinates is located in the point $\mu^*=0$, $\Delta=0$. The abscissa axis is along the
line $\mu^*=0$. The ordinate axis is along the rectilinear diameter $\Delta=0$.}
	\label{fig:mu_w}
\end{figure}

The quantity $\Xi_1$ is the partition function~\cite{Yukh_Kol_Idz_13}, which includes all
interaction effects, with the exception of the effects connected with the formation of a new
phase, i.e., the liquid phase within the maternal sphere of gas in the volume $V_0$:
\be
\label{eq7}
\Xi_1 = \Xi_\textrm{g} \exp\left[ - \beta \left(F_\textrm{CR} + F_\textrm{IGR}\right)\right].
\ee
Here, $\Xi_\textrm{g}$ is the partition function in which the effects appearing after integration
over the collective variables $\rho_{\bf k}$ for $k > B$ are taken into account, where $B$ is the point of the
first zero of the Fourier transform of the attraction potential;
the integration is performed with the Gaussian density measure; $F_\textrm{CR} + F_\textrm{IGR}$
are the free energies arising after integration over  $\rho_{\bf k}$ with $0 < {\bf k} \leqslant B$: $F_\textrm{CR}$
in the critical regime, where the renormalization group (RG) symmetry between the block-Hamiltonians exists and
${2\pi n_{\tau}}/{a_{n_\tau}} \leqslant {\bf k} \leqslant B$, $n_{\tau}$ is the number and $a_{n_\tau}$
is the periods of the block-lattice, on which,  given $\tau$, the renorm-group symmetry
between the coefficients of the block-Hamiltonians terminates; the integration is
performed in the quartic density measure~\cite{Yukh_87,Yukh_nuovo89,Mic_Theory_01}; $F_\textrm{IGR}$
appears as a result of the integration  over $\rho_{\bf k}$ performed here  in an inverse-Gaussian
regime~\cite{Yukh_87} for $0 < {\bf k} \leqslant {2\pi n_{\tau}}/{a_{n_\tau}}$. We suppose that all
expressions in $\Xi_1$ are known, and that all of them are fixed quantities within the temperature
region $\tau < 0.01$. In this work, we consider only one part, $\Xi_{\rho_0}$, of the $\Xi$. It is
connected with the density fluctuations of the number of particles inside of the sphere of the radius  $R_0$.
The phenomena of the phase transition of the first order at $\tau < 0$ are described by the expression $\Xi_{\rho_0}$.
The chemical potential  $\mu$ of the whole system is only within $\Xi_{\rho_0}$.
As a result of calculations carried out in~\cite{Yukh_Kol_Idz_13,Yukh_Idz_Kol_95},
the initial form of the partition function $\Xi_{\rho_0}$ for integration over $\rho_0$ is as follows:
\be
\label{eq8}
\Xi_{\rho_0} = \exp \Bigl[ \mu^*(1 - \Delta) N\Bigr] \int \exp [NE(\rho_0)] \rd\rho_0\,,
\ee
where $\mu^*$ is the generalized chemical potential of the system\footnote{While investigating
the characteristics of the integrand, the origin of the coordinates is transferred to
the point $\Delta=0$, $\mu^*=0$, and thus the change of $\rho_0$ may accept both
positive and negative values (see figure~\ref{fig:mu_w}).},
\be
\label{eq9}
\mu^* = \beta (\mu - \mu_0) + \xi - | \alpha(0)| (1 - \Delta),
\ee
$\mu_0$ is the chemical potential of the RS,
\be
\label{eq10}
\xi = \frac{{\mathfrak{M}}_3}{|{\mathfrak{M}}_4|}\,,
\ee
\be
\label{eq11}
\Delta = - \left(\xi {\mathfrak{M}}_2 + \frac13 {\mathfrak{M}}_3 \xi^2\right),
\ee
${\mathfrak{M}}_2$, ${\mathfrak{M}}_3$, ${\mathfrak{M}}_4$ are the cumulants of the Jacobian of transition~\cite{Yukh_87,Yukh_nuovo89};
at small densities, $\Delta$ is a linear function  of density $\eta$, see figure~\ref{fig:Delta_crit}.
\be
\label{eq12}
\alpha(0) = \left( \frac{N}{V} \frac{\tilde \Phi(\bf k)}{\Theta} \right),
\ee
where $\tilde \Phi(\bf k)$ is the Fourier transform of the attraction potential;
\be
\label{eq13}
E(\rho_0) = \mu^* \rho_0 + D \rho_0^2 - G\rho_0^4,
\ee
\be
\label{eq14}
D = D_0 |\tau|^{2\nu}, \qquad G = G_0|\tau|^{\nu}
\ee
are coefficients of the limiting block-structure Hamiltonian $H_{n_\tau}$ at the
variable $\rho_0$, $D_0 \simeq 1.19$; $G_0 = 1.67$; $\nu = \ln s^*/\ln E_1 = 0.605$ is
the correlation length critical exponent, $E_1$ is the greater of two eigenvalues of the
matrix for linearized recurrent relations of Wilson's
type~\cite{Yukh_87,Yukh_nuovo89,Mic_Theory_01,widom:3898,Kad_66,Wil_71a,Wil_71b}
for coefficients of block-structure Hamiltonians, $s^*$ is the optimal parameter
in the scale transformations $s^* = 3.58\ldots$, $(s^*) < E_1 < (s^*)^2$. We work in a
critical region of temperatures and densities\footnote{``critical'' means that within
this region there exists the RG symmetry.}, $\tau < \tau^*$, $\eta < \eta^*$. For the limit
interval of the temperature $\tau^*$ we have the expression:~\cite{Yukh_nuovo89}
\be
\label{eq15}
\left[ \frac{\tilde C_1 \tau^*}{r^* + \beta \tilde \Phi (0) \eta \frac{6}{\pi\sigma^3}} \right]^\nu =
\left[ 1 - \frac{a_2}{\beta \tilde \Phi (0) \eta \frac{6}{\pi\sigma^3}} \right]^{1/2}.
\ee
Here, $r^*$ is one of the fixed-point coordinates, $a_2$ is a coefficient in the
Jacobian of transition. On both sides of~(\ref{eq15}) there are inverse radia of correlation
in the quartic and in Gaussian density measure, correspondingly~\cite{Yukh_nuovo89}.

For the limiting point of the interval of densities we have:
\be
\label{eq16}
|\eta^* - \eta_\textrm{c}| = \eta_\textrm{c} \sqrt{\frac{D_0}{G_0}}(\tau^*)^{\nu/2},
\ee
where $\eta = \frac{N}{V} \frac{\pi \sigma^3}{6}$, $\sigma$ is a diameter
of a particle (hard core), $\eta^{*}$ is the limiting value of density for
the critical regime.

In general, coefficients $D_0$ and $G_0$ in~(\ref{eq13}) depend on density.
In this work we take them as $\eta = \eta_\textrm{c}$, where $\eta_\textrm{c} = 0.13044$ is the
critical density determined from~(\ref{eq11}) by the condition $\Delta =0$.
The Hamiltonian $E(\rho_0)$ in~(\ref{eq8}) is obtained as a result of
integration in~(\ref{eq5}) over all variables $\rho_{\bf k}$, with the
exception of $\rho_0$\footnote{The variable $\rho_0$ is introduced in $\Xi$ in~(\ref{eq5})
as a result of integration over $d\Gamma_N$ and summation over $N$ of the
expression $\delta(\rho_0 - {N}/{\sqrt{\langle N \rangle}})$,
where $N$ is the independent variable of the number of particles. In order
to integrate over $\rho_0$ in~(\ref{eq8}) we surely have to take the grand partition function.}.

The chemical potential $\mu$ appears only in $\Xi_{\rho_0}$. Thus,
the variable $\rho_0$ is connected with the density and with the distribution of the number of particles.

By definition
\be
\label{eq17}
\ln \Xi = \frac{PV}{\Theta} \quad \quad {\textrm{and}} \quad \quad
\frac{\partial \ln \Xi}{\partial (\beta\mu)} = \frac{\partial \ln \Xi_{\rho_0}}{\partial \mu^*} = \langle N \rangle .
\ee
Based on these two expressions we get the equation of state:
\be
\label{eq18}
P = p(\tau, \eta),
\ee
connecting such variables as pressure, temperature, density, $p$, $\tau$, $\eta$.
The curve of equations~(\ref{eq17}) is an envelope curve for ${pV}/{\theta} = \ln \Xi$,
the latter depends on $p$, $\tau$, $\eta$, $\mu$.

We work within the framework of the equilibrium thermodynamics. The number $N$ is large.
We use the presence of the thermodynamic limit:
\be
\label{eq19}
N \to \infty, \qquad V \to \infty; \qquad \frac{N}{V} = {\rm const}.
\ee
In this connection, when we integrate over $\rho_0$ in ~(\ref{eq8}) we use the
steepest descent method. The integration in~(\ref{eq8}) was carried out and
discussed in~\cite{Yukh_Kol_Idz_13,Yukh_Idz_Kol_95}. We give here only the final conclusions.

In the integral $\int \exp [NE(\rho_0)]\rd\rho_0$ in~(\ref{eq8}) we looked for
the maximum of the  subintegral function.

The derivative
\be
\label{eq20}
\frac{\partial E (\rho_0)}{\partial \rho_0} = \mu^* + 2D\rho_0 - 4 G\rho_0^3 = 0
\ee
gives us the cubic equations:
\be
\label{eq21}
\rho_0^3 + V\rho_0 + W = 0, \qquad {\textrm{where}}  \qquad V = -\frac12 \frac{D}{G}\,, \qquad W = - \frac{\mu^*}{4G} \,,
\ee
on condition of $\ddot{E}(\rho_0) < 0$:
\be
\label{eq22}
2D - 12G \rho_0^2 < 0 \qquad {\textrm{or}} \qquad 3(\rho_0^{\rm max})^2 + V > 0.
\ee
The equation~(\ref{eq21}) has three roots, depending on the sign of the discriminant $Q$:
\be
\label{eq23}
Q = \frac{W^2}{4} + \frac{V^3}{27}\,.
\ee
Here, the first addend is positive, and the second is always negative.
Therefore, the discriminant $Q$ of equation~(\ref{eq21}) may be positive,
equal to zero and negative.  The case of $Q=0$ separates the roots.
At $Q=0$ $\left( {W}/{2}\right)^2 = - \left( {V}/{3}\right)^3$, we
have three real roots, among them only the root $\rho_0^{\rm max} = 2\sqrt[3]{-W/2} = \sqrt[3]{\mu^*/G}$
meets the requirement~(\ref{eq22})\footnote{Two other roots,
$\rho_2$ and $\rho_3 = - \frac12 \sqrt[3]{\mu^*/G}$ do not obey~(\ref{eq22}).}.
From equation~(\ref{eq23}) at $Q=0$ we get
\be
\label{eq24}
\mu^* = \pm G \left(\sqrt{\frac23 \frac{D}{G}}\right)^{3} \qquad \textrm{and thus} \qquad
\rho_0^{\rm max} = \pm \sqrt{\frac23 \frac{D}{G}}\,.
\ee
The roots~(\ref{eq24}) are the vertex of the rectangle on the plane $\mu^*$, $\rho_0$. Let us introduce the notation
\be
\label{eq25}
a = Gb^3, \quad b = \sqrt{\frac23 \frac{D}{G}}\,.
\ee

Out of the rectangle we have the region  $Q > 0$, inside $Q < 0$. In the case
of $Q > 0$, equation~(\ref{eq20}), or~(\ref{eq21}), has only one root~\cite{Yukh_Kol_Idz_13}
\be
\label{eq26}
\rho_0^{\rm max} = \rho_1 = q\left( \frac{\mu^*}{G} \right)^{1/3}  , \qquad \mu^* = \pm G \left( \frac{\rho_1}{q}\right)^3,
\ee
where
\bea
&& q = 4^{-1/3}\left\{ 1 + \left( \frac{|\gamma|}{4} \right)^{1/3} + \ldots \right\}, \qquad
\gamma = \left( - \frac{V}{3}\right)^3 \Big/ \left( \frac{W}{2}\right)^2 < 1, \nonumber\\
&& q = 4^{-1/3} \qquad\qquad \, {\textrm{at}} \qquad \gamma = 0, \nonumber\\
&& q = 1 \qquad\qquad\qquad {\textrm{at}} \qquad \gamma = 1, \nonumber
\eea
which corresponds to~(\ref{eq24}) at $Q=0$.

In the case of $Q < 0$, we have three real roots:
\be
\label{eq27}
\mu^* = a \cos \gamma, \qquad {\textrm{and}} \qquad  \rho_{1,2,3} = b \cos \frac{\gamma + 2k\pi}{3}\,.
\ee
$n = 0, 1, 2$. Only the roots
\be
\label{eq28}
\rho_{01} = b \cos \frac{\varphi}{3} \quad {\textrm{and}} \quad  \rho_{02} = b \cos \frac{\varphi + 2\pi}{3}
\ee
meet the requiremets~(\ref{eq22}). The principal maximum of $\exp [NE(\rho_0)]$
for the root $\rho_{02}$ takes place for $\frac{\pi}{2} \leqslant \varphi \leqslant \pi$
(the thick line in figure~\ref{fig:mu_w}), and for the root $\rho_{01}$,
when $0 \leqslant \varphi \leqslant \frac{\pi}{2}$. The secondary maxima of $\exp [NE(\rho_0)]$
exist for $\rho_{02}$ in the region $0 \leqslant \varphi < \frac{\pi}{2}$ and for  $\rho_{01}$
in the region $\frac{\pi}{2} \leqslant \varphi \leqslant \pi$.

The curve for the principal maxima of the generalized chemical potential $\mu^*$
has a jump in $\rho_{0}$ from $-d$ to $+d$ on the axis $\mu^* = 0$.

A smooth transition of the solution $\rho_{1}$ into $\rho_{02}$ at the point
$(-b, -a)$ and $\rho_{1}$ into $\rho_{01}$ at the point $(b, a)$ takes place.

Herein below we shall consider an isothermal process of a quasistatic pressure on the
gas system at $\tau = {\rm const}$, $\tau < 0$, $|\tau| < \tau^*$.

\section{Monophasic states of liquid and gas}
\label{sec:monostates}
The solution $\rho_{\rm max}^0 = \rho_1$ in~(\ref{eq26}), describes the values of
densities of absolute maxima for the function $\exp [NE(\rho)]$ in integral~(\ref{eq8})
for $\Xi_{\rho_0}$. Considering $N$ as a very big number, according to the steepest-descent
method, we write:
\be
\label{eq1.1}
\ln \Xi_{\rho_0} = N (1 - \Delta) \mu^* + NE(\rho_1).
\ee
Eliminating $\mu^*$ by means of~(\ref{eq17})
\be
\label{eq1.2}
\frac{\partial \ln \Xi_{\rho_0}}{\partial \mu^*} = N
\ee
one gets $(1 - \Delta)N + N \rho_1 = N$, and
\be
\label{eq1.3}
\rho_1 = \Delta.
\ee
Then, in correspondence with (\ref{eq20})
\be
\label{eq1.4}
\mu^* = \mu^*(\Delta) = -2D\Delta + 4G\Delta^3 \qquad {\textrm{and}} \qquad
E(\rho_0^{\rm max}) = \mu^* \Delta + D\Delta^2 - G\Delta^4 .
\ee

Relationships~(\ref{eq20})--(\ref{eq1.2}) together with~(\ref{eq5})--(\ref{eq8})
give us the equation of state
\be
\label{eq1.5}
PV = P_0 V + P_1V + P_{\rho_0}V,
\ee
where
\be
\label{eq1.6}
P_0V = \Theta \ln \Xi_0
\ee
--- contribution from the reference system,
\be
\label{eq1.7}
P_1V = \Theta \ln \Xi_1
\ee
is the contribution from the region $k \geqslant B$ and from the integrals
over $\rho_{\bf k}$ for $0 < {\bf k} \leqslant B$. The latter includes integrals
in the critical regime and in the inverse Gaussian regime, and the free energy
of the RS up to the cumulant ${\mathfrak{M}}_4$.

Finally,
\be
\label{eq1.8}
P_{\rho_0}V = \Theta \ln \Xi_{\rho_0}\,,
\ee
\be
\label{eq1.9}
\ln \Xi_{\rho_0} = N {\cal E}(\Delta),
\ee
where in accordance with~(\ref{eq1.1})--(\ref{eq1.4})
\begin{align}
\label{eq1.10}
{\cal E}(\Delta) &= \mu^*(\Delta) + D\Delta^2 - G\Delta^4,\\
\label{eq1.11}
\mu^{*}(\Delta) &= - 2 D\Delta + 4G\Delta^3,\\
\label{eq1.12}
P_{\rho_0}V &= \Theta N \left( -2D\Delta + 4G\Delta^3 + D\Delta^2 - G\Delta^4 \right).
\end{align}
The sum of~(\ref{eq1.6}), (\ref{eq1.7}) and~(\ref{eq1.12}) in~(\ref{eq1.5})
presents the gas-state isotherm when $\Delta \leqslant -b$, and the liquid-state isotherm when $\Delta \geqslant b$.

Our main task is to consider events in the region $|\Delta| \leqslant b$. In this region,
the discriminant~(\ref{eq23}) of equation~(\ref{eq21}) is negative. Note that we
restrict ourselves to $\tau < 0$, $|\tau| < \tau^* \simeq 0,02$ and
\be
\label{eq1.13}
|\eta - \eta_\textrm{c}| < \eta_\textrm{c} \sqrt{\frac{D_0}{G_0}}(\tau^*)^{\nu/2}.
\ee
By definition, the coordinates of the critical point are $\tau = 0$, $\Delta (\eta_\textrm{c}) = 0$, $\eta_\textrm{c} = 0.13044$.

The expression for $\Delta$ is given in~(\ref{eq11}). The quantity $\Delta$ is
also determined  in~(\ref{eq1.3}). Here, for $\rho_1$ we have its value~(\ref{eq26}),
the quantities $D$ and $G$ being connected with the attractive forces. Thus, owing
to expression~(\ref{eq1.2}), both the connection of the effects of the long-range and
of the short-range (of the RS) interactions takes place.

\begin{figure}[!t]
	\centerline{
\includegraphics[angle=0,width=0.55\textwidth]{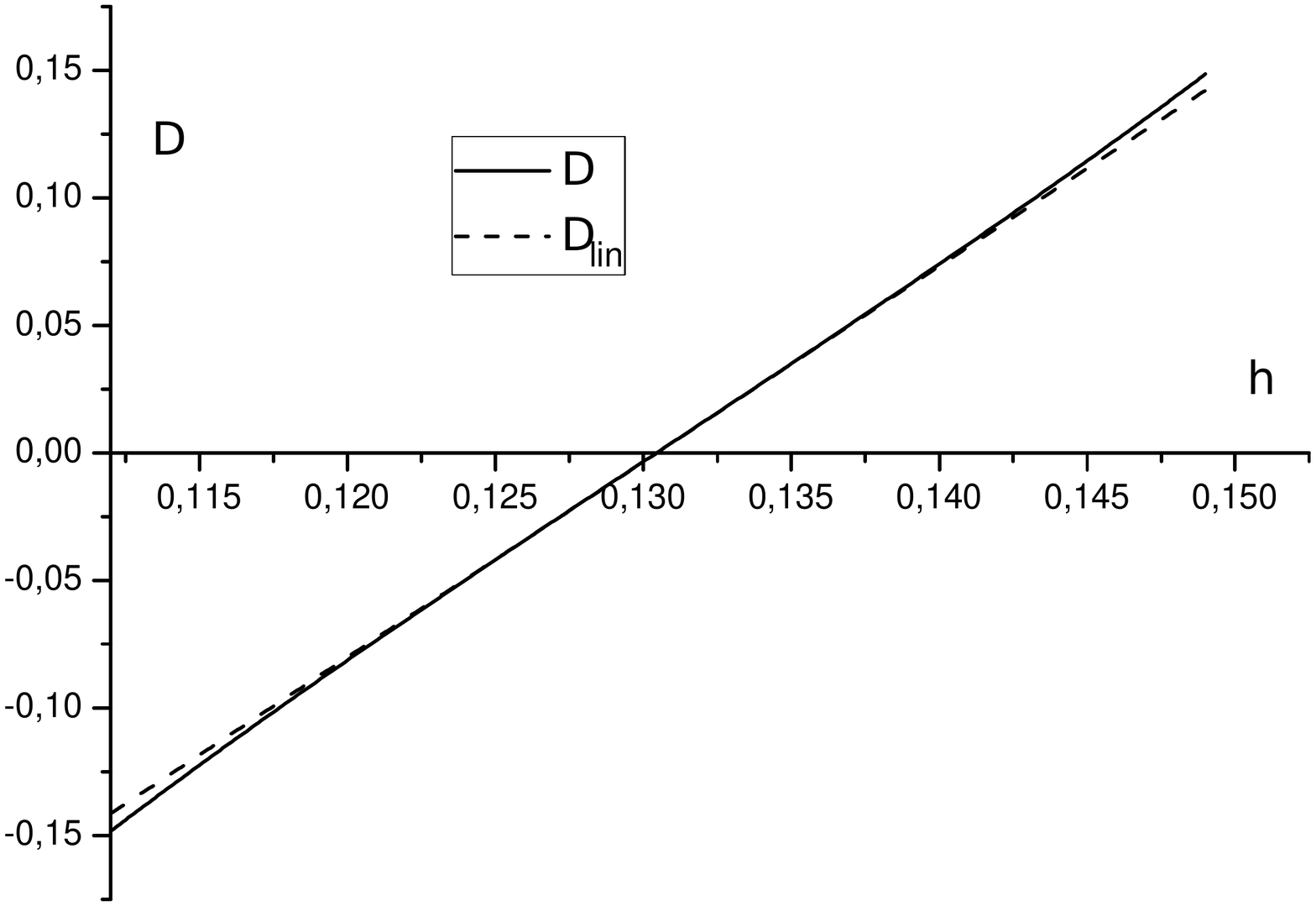}
}
	\caption{The region of the linear approximation for $\Delta$ as a function of
the density and vice versa. Our discussion concerns the critical region of densities
[see equation~(\ref{eq16})] $|\eta - \eta_\textrm{c}| \leqslant 0.02$ $\Delta = - ({\mathfrak{M}}_2\xi
+ \frac13 {\mathfrak{M}}_3\xi^2)$, $\eta = ({N}/{V})({\pi \sigma^3}/{6})$, $\eta_\textrm{c} = 0.13044$.}
	\label{fig:Delta_crit}
\end{figure}

In the critical region, we can also use for  $\Delta$ the following expansion:
\be
\label{eq1.14}
\Delta(\eta) = \Delta(\eta_\textrm{c}) - \left( \frac{\rd \Delta}{\rd\eta} \right)_{\eta_\textrm{c}}(\eta - \eta_\textrm{c}).
\ee
Here, the following expressions hold $\left( {\rd \Delta}/{\rd\eta} \right)_{\eta = \eta_\textrm{c}} = {1}/{\eta_\textrm{c}}$.
According to the definition of the critical point $\Delta |_{\eta = \eta_\textrm{c}} = 0$, and for~(\ref{eq1.14}) we get (see figure~\ref{fig:Delta_crit}):
\be
\label{eq1.15}
\Delta = \frac{1}{\eta_\textrm{c}} (\eta - \eta_\textrm{c})
\ee
and
\be
\eta = \eta_\textrm{c} \Delta + \eta_\textrm{c}\,.
\ee

It follows that for gaseous state $\rho_1 \leqslant -b$ and for liquid state $\rho_1 \geqslant b$ in~(\ref{eq1.15}),
one has some restrictions on values of $\Delta$ and $\eta$. Besides, we should take
into account the restrictions on density in~(\ref{eq1.13}), as well as those connected with
cumulants ${\mathfrak{M}}_4$~\cite{Mic_Theory_01,Yukh_Idz_Kol_95}.

For $\mu^*$ we have expressions~(\ref{eq1.4}) and~(\ref{eq9}). The latter includes the
chemical potential of the reference system, and~(\ref{eq1.5}) has a contribution into the pressure of the whole system.

\section{On the reference system}
\label{sec:ref_system}
Let us turn back to formula~(\ref{eq5}) for the grand partition function $\Xi$.
The logarithm of $\Xi$ gives the whole equation of state accounting for external shell,
thus accordingt to~(\ref{eq1}) $\ln \Xi = P_{\textrm{sph}}V_{\textrm{sph}} - \alpha S_{\textrm{sph}}$,
where according to~(\ref{eq2}), $P_{\textrm{sph}} = p + \delta P_0$.
In this section, we omit the terms connected with the existence of the external shell assuming
$\ln \Xi = {PV}/{\theta}$. In accordance with~(\ref{eq5})
\be
\label{eq2.1}
\ln \Xi = \ln \Xi_0 + \ln \Xi_1 + \ln \Xi_{\rho_0}\,.
\ee
Our aim is to separate in the right-hand side the terms belonging only to the RS.
At first it is $\ln \Xi_0 = {P_0V}/{\theta}$, where $P_0$ is the pressure of the RS.
Then, from~(\ref{eq1.9}),  (\ref{eq1.10}) and (\ref{eq9}) for $\ln \Xi_{\rho_0}$ we may write:
\be
\label{eq2.2}
\ln \Xi_{\rho_0} = N \Bigl[ \beta (\mu - \mu_0) + \xi - |\alpha(0)|(1-\Delta) \Bigr] + N(D\Delta^2 - G\Delta^4).
\ee
Thus, in the right-hand side we have two addends connected with RS, namely, $N(\xi - \beta\mu_0)$.
They do not disappear, when excluding the attractive interaction, taking $\alpha = 0$.
Of course, in the right-hand side of~(\ref{eq2.1}) we should exclude some additional
expressions, completing together the free energy $F_0$ of the RS. We are talking about
the factor\footnote{It is shown in~\cite{Yukh_Idz_Kol_95}, $Q(-\xi, a_2, a_4)
= \left[ \int_{-\infty}^{\infty}\exp \left( -\xi \rho_l - \frac{a_2}{2}\rho_l^2
- \frac{a_4}{4!}\rho_l^4 \right) \rd\rho_l \right]^{N_\textrm{B}}$, where $a_2, a_4$ are
the coefficients in the transition Jacobian into the phase space of collective variables.}
$\ln Q^{-1} (-\xi, a_2, a_4)$ that forms a part in the expression for $\Xi_1$.
Collecting together all terms remaining in the expression~(\ref{eq2.1}) at $\alpha = 0$,
we can write~(\ref{eq2.1}) in the form:
\be
\label{eq2.3}
\frac{PV}{\Theta} - \frac{P_0V}{\Theta} = N \{ \mu - \mu_0 \} - (F - F_0),
\ee
where ${P_0V}/{\Theta} = \ln \Xi_0$, and
\bea
\label{eq2.4}
 F - F_0 &=& N \Bigl[ -|\alpha(0)|(1 - \Delta) + N \left(D\Delta^2 - G\Delta^4\right) \Bigr] + \nonumber\\
&& + \ln \Xi_\textrm{g} - \beta (F_\textrm{CR} + F_\textrm{IGR}) - N \left[ -\xi + \frac{N_B}{N} \ln Q (\xi, a_2, a_4) \right].
\eea
As a result, we may write $F_0$ in the form
\be
\label{eq2.5}
\frac{F_0}{\Theta} = N \xi - N_\textrm{B} \ln Q (\xi, a_2, a_4).
\ee

Herein below in our computations, in the critical region we shall consider the quantities entering (\ref{eq2.4}) as the known
quantities. We have presented here the expressions for $F_0$ because we say that the pressure
$P_0$, and the chemical potential $\mu_0$ ought to be determined from the expression for $F_0$.
The latter is obtained  here using the quartic density measure while calculating the Jacobian.
Thus, for the chemical potential of the RS we shall take the expression
\be
\label{eq2.6}
\mu_0 = \frac{\partial F_0}{\partial N} \simeq \xi .
\ee
The extraction of the RS while calculating the equation of state according to~(\ref{eq2.3}),
will not be considered here, and will be the subject of another research.

\section{The field of the phase transition}
\label{sec:phase_trans}
We consider the partition function~(\ref{eq5}) in the case of negative values of the
discriminant~(\ref{eq23}) of equation~(\ref{eq20}). The part of the expression~(\ref{eq5})
will be analysed in detail, which is connected with the chemical potential via $\Xi_{\rho_0}$
according to~(\ref{eq17}). For $\Xi_{\rho_0}$, we have expressions~(\ref{eq8})--(\ref{eq13}).
We shall remember here the main form
\be
\label{eq3.1}
\Xi_{\rho_0} = \exp \Bigl[ N\mu^* (1 - \Delta) \Bigr] \int\limits_{-\infty}^{\infty} \exp \Bigl[ N E(\rho) \Bigr] \rd\rho .
\ee
In the region $Q < 0$, equation~(\ref{eq20}) for extremum of $\exp N E(\rho)$ has the roots of~(\ref{eq27}).

We shall investigate the equation~(\ref{eq17}) for the envelope curve in the case of $Q < 0$.

Let us consider more in detail the second of the two equations
\be
\label{eq3.2}
\frac{\partial \ln \Xi_{\rho_0}}{\partial \mu^*} - N = 0.
\ee
Taking into account~(\ref{eq3.1}), one has
\[
N(1 - \Delta) + \frac{\int\limits_{-\infty}^{\infty} N\rho \exp  (N E(\rho)) \rd\rho}{\int\limits_{-\infty}^{\infty} \exp ( N E(\rho)) \rd\rho} = N,
\]
or
\be
\label{eq3.3}
\frac{\int\limits_{-\infty}^{\infty} \rho \exp  [N E(\rho)] \rd\rho}{\int\limits_{-\infty}^{\infty} \exp [N E(\rho)] \rd\rho} = \Delta .
\ee
On the right we have $\Delta = \Delta (\eta)$, given in~(\ref{eq11}),
where the cumulants of the RS are present. On the left-hand side of equation~(\ref{eq3.3})
there are the magnitudes connected with the long-range effects of the Van der Waals forces.
This important equation connects together the long-range effects accumulated in the left-hand side, and the short-range effects accumulated in $\Delta$. As we remember, long-range
forces are described in the collective variables phase space and the short-range ones in the
Cartesian-phase space.

We devide the integral $\int_{-\infty}^{\infty} \rho \exp [N E(\rho)] \rd\rho$ into two parts
\[
\int\limits_{-\infty}^{\infty} \rho \, \exp [N E(\rho)] \rd\rho = \int\limits_{-\infty}^{0} \rho \, \exp [N E(\rho)] \rd\rho
+ \int\limits_{0}^{\infty} \rho \, \exp [N E(\rho)] \rd\rho .
\]
In the first one, along with figure~\ref{fig:mu_w}, the maximum of the subintegral
function $\exp [N E(\rho)]$ takes place for $\rho = \rho_{02} = b\cos \frac{\varphi + 2\pi}{3}$,
$0 \leqslant \varphi \leqslant \pi$. This is the domain of the gas state of the matter.
In the second term, the maximum for $\exp [N E(\rho)]$ takes place for $\rho = \rho_{01} = b\cos \frac{\varphi}{3}$~--- this is the liquid domain.

In the whole integral~(\ref{eq3.3}), the principal maximum for the gas-phase trips together with the secondary maximum for liquid, and vice-versa the principal
maximum for the liquid phase trips together with the secondary maximum for the gas.

Let us go to a more detailed consideration of expressions~(\ref{eq3.3}).

So, we carry out the integrals in~(\ref{eq3.3}), and denote by
\bea
\label{eq3.4}
 {\cal I} = \int\limits_{-\infty}^{\infty} \rho \, \exp [ N E(\rho) ] \rd\rho , \qquad {\textrm{and by}}  \qquad
{\cal K} = \int\limits_{-\infty}^{\infty} \exp [N E(\rho) ] \rd\rho .
\eea
Taking into account that the maxima of the subintegral function go along
the lines $\rho = \rho_{02}$ and $\rho = \rho_{01}$ and that the curve $\rho_{02}$,
where $-b \leqslant \rho_{02} \leqslant -b/2$, is connected with the gas-phase, and the curve
$\rho_{01}$, where $b/2 \leqslant \rho_{01} \leqslant b$, is connected with the liquid phase, we shall write
down integrals~(\ref{eq3.4}) as the sums:
\bea
\label{eq3.5}
&& {\cal I} = {\cal I}_{\,\textrm{g}} + {\cal I}_{\,\textrm{l}} \qquad {\textrm{and}} \qquad
{\cal K} = {\cal K}_{\textrm{g}} + {\cal K}_{\textrm{l}},
\eea
where
\begin{align}
&{\cal I}_{\,\textrm{g}} =  \int\limits_{-\infty}^0 \rho \, \exp [N E(\rho)] \rd\rho ,
& &{\cal I}_{\,\textrm{l}} = \int\limits_{0}^{\infty} \rho \, \exp [ N E(\rho) ] \rd\rho ,\\ %
& {\cal K}_{\textrm{g}} =  \int\limits_{-\infty}^0  \exp [N E(\rho)] \rd\rho ,
& & {\cal K}_{\textrm{l}} =  \int\limits_{0}^{\infty} \exp [ N E(\rho)] \rd\rho .
\end{align}
All of them are of the same type, and we shall calculate ${\cal I}_{\,\textrm{g}}$ precisely.
We expand the function $E(\rho)$ at the maximum point $\rho = \rho_{02} = b \cos \frac{\varphi + 2\pi}{3}$. Then,
\be
\label{eq3.6}
{\cal I}_{\,\textrm{g}} = \exp [ N E(\rho_{02})] ({\cal I}_{11}\rho_{02} + {\cal I}_{12}),
\ee
where
\bea
 {\cal I}_{11} &=& \int\limits_{-\infty}^0  \exp \left[ - \frac{N}{2} | \ddot{E} (\rho_{02}) | ( \rho - \rho_{02} )^2 \right] \rd\rho = \nonumber\\
& = &\left\{  \frac{\sqrt{\pi}}{2} \left[ \frac{N}{2} | \ddot{E} (\rho_{02}) | \right]^{-1/2} +
\int\limits_0^{|\rho_{02}|} \exp \left[ -\frac{N}{2} | \ddot{E} (\rho_{02}) | x^2\right] \rd x  \right\}, \nonumber\\
 {\cal I}_{12} &=& \int\limits_{-\infty}^0 (\rho - \rho_{02}) \exp \left[ - \frac{N}{2} | \ddot{E} (\rho_{02}) | ( \rho - \rho_{02} )^2\right] \rd\rho_{02} = \nonumber\\
&=& \frac{1}{|N \ddot{E}(\rho_{02})|} \exp - \frac12 N | \ddot{E}(\rho_{02}) |  \rho_{02}^2\,. \nonumber
\eea

For the second integral in ${\cal I}_{11}$, we introduce the probability integral
and get\footnote{While calculating ${\cal I}_{11}$, ${\cal I}_{12}$ and ${\cal K}$
we used $\exp N \left[ E(\rho_{02}) - \frac12 | \ddot{E}(\rho_{02}) \rho_{02}^2 \right] \simeq \exp [NE (0)]$, $E(0) = 0$.}
\bea
&& \int\limits_{0}^{|\rho_{02}|}  \exp \left[ - \frac{N}{2} \ddot{E} (\rho_{02}) x^2\right] \rd x =
\frac{\sqrt{\pi}}{2} ({\rm erfc} \, z) \frac{1}{\sqrt{\frac{N}{2} |\ddot{E}  (\rho_{02})|}}\,, \nonumber
\eea
\be
 z = \sqrt{\frac{N}{2} |  \ddot{E} (\rho_{02})|} |\rho_{02}|\,, \qquad
{\rm erfc} \, z = \frac{1}{\sqrt{\pi}} \frac{1}{z} \re^{-z^2} \left[ 1 + \sum\limits_{m=1}^{\infty} (-1)^m
\frac{1.3 \dots (2m -1)}{(2z^2)^m} \right]. \nonumber
\ee
Substituting into~(\ref{eq3.6}) and gathering together all the terms after
shortening in the limit of big $N$, we find
\be
\label{eq3.7}
{\cal I}_{\,\textrm{g}} = \rho_{02} \re^{NE(\rho_{02})}
\frac{\sqrt{\pi}}{\sqrt{\frac{N}{2} | \ddot{E} (\rho_{02})|}}\,.
\ee
The integral ${\cal I}_{\,\textrm{l}}$ when calculated undergoes the same
transformations and we obtain
\be
\label{eq3.8}
{\cal I}_{\,\textrm{l}} = \rho_{01} \re^{NE(\rho_{01})}
\frac{\sqrt{\pi}}{\sqrt{\frac{N}{2} | \ddot{E}  (\rho_{01})|}}\,.
\ee
The integrals ${\cal K}_{\textrm{g}}$ and ${\cal K}_{\textrm{l}}$ are actually
calculated while calculating  ${\cal I}_{11}$. For ${\cal K}_{\textrm{g}}$ and ${\cal K}_{\textrm{l}}$, we have:
\bea
\label{eq3.9}
&& {\cal K}_{\textrm{g}} =  \exp [NE(\rho_{02})]
\frac{\sqrt{\pi}}{\sqrt{\frac{N}{2} |\ddot{E} (\rho_{02})|}} \,, \nonumber\\
&& {\cal K}_{\textrm{l}} = \exp [NE(\rho_{01})]
\frac{\sqrt{\pi}}{\sqrt{\frac{N}{2} |\ddot{E} (\rho_{01})|}} \,.
\eea
After all these calculations, we return to expressions~(\ref{eq3.2}), ~(\ref{eq3.3}),
that together with~(\ref{eq3.1}) describe the envelope-curve equation, or, in other
words, to the equation of state in the region of $Q < 0$. Expression~(\ref{eq3.3}) now has the form:
\bea
\label{eq3.10}
&& \Delta = \frac{{\cal I}_{\,\textrm{g}} + {\cal I}_{\,\textrm{l}}}
{{\cal K}_{\textrm{g}} + {\cal K}_{\textrm{l}}}\,, \qquad {\textrm{or explicitly}} \qquad \Delta = \rho_{02} w_{\textrm{g}} +\rho_{01} w_{\textrm{l}}\,,
\eea
where
\bea
&& w_{\textrm{g}} = \frac{c_{\textrm{g}} \exp [NE(\rho_{02})]}
{c_{\textrm{g}} \exp [NE(\rho_{02})] + c_{\textrm{l}} \exp [NE(\rho_{01})]} = \frac{{\cal K}_{\textrm{g}}} {{\cal K}_{\textrm{g}} + {\cal K}_{\textrm{l}}}\,, \nonumber\\[1ex]
&& w_{\textrm{l}} = \frac{c_{\textrm{l}} \exp [NE(\rho_{01})]}
{c_{\textrm{g}} \exp [NE(\rho_{02})] + c_{\textrm{l}} \exp [NE(\rho_{01})]} = \frac{{\cal K}_{\textrm{l}}} {{\cal K}_{\textrm{g}} + {\cal K}_{\textrm{l}}}\,, \nonumber\\[1ex]
&& c_{\textrm{g}} = (|\ddot{E} (\rho_{02})|)^{-1/2}, \qquad
c_{\textrm{l}} = |\ddot{E} (\rho_{01})|^{-1/2}. \nonumber
\eea
Here,  the functions $w_{\textrm{g}}$ and $w_{\textrm{l}}$ play the role of the
probabilities of the gas or of the liquid state, accordingly.

\bea
\label{eq3.11}
w_{\textrm{l}} = \left\{ \begin{array}{ccl}
	1 & \textrm{at} & \varphi = 0 \,,  \\
	\frac12 & \textrm{at} & \varphi = \frac{\pi}{2}\,,    \\
	0 & \textrm{at} &  \varphi = \pi\,;
\end{array}
\right.
\qquad
w_{\textrm{g}} = \left\{ \begin{array}{ccl}
	0 & \textrm{at} & \varphi = 0,   \\
	\frac12 & \textrm{at} & \varphi = \frac{\pi}{2},    \\
	1 & \textrm{at} &  \varphi = \pi.
\end{array}
\right.    \nonumber
\eea

As we see,
\be
\label{eq3.12}
w_{\textrm{g}} + w_{\textrm{l}} = 1.
\ee
Herein,
\be
\rho_{02}=b\cos{\frac{\varphi+2\pi}{3}} = \left\{
\begin{array}{ccl}
	-b,& \varphi = \pi, \\
	-d,& \varphi = \frac{\pi}{2}, \\
	-\frac{b}{2}, & \varphi = 0.
\end{array}
\right.
\ee
\be
\rho_{01}=b\cos{\frac{\varphi}{3}} = \left\{
\begin{array}{ccl}
	\frac{b}{2}, & \varphi  = \pi, \\
	d, & \varphi  = \frac{\pi}{2}, \\
	b, & \varphi  = 0.
\end{array}
\right.
\ee
For $\Delta$, we have:
\be
\label{eq3.13}
\Delta = \Delta_\textrm{g} + \Delta_\textrm{l}\,,
\ee
\be
\label{eq3.13a}
\Delta_\textrm{g}= w_\textrm{g}\rho_{02} = \left\{
\begin{array}{ccl}
	-b,& \varphi  = \pi, \\
	-d/2,& \varphi  = \frac{\pi}{2}, \\
	0, & \varphi  = 0.
\end{array}
\right.
\ee
\be
\label{eq3.13b}
\Delta_\textrm{l}= w_\textrm{l}\rho_{01} = \left\{
\begin{array}{ccl}
	0,& \varphi  = \pi, \\
	d/2,& \varphi  = \frac{\pi}{2}, \\
	b, & \varphi  = 0.
\end{array}
\right.
\ee
Here, $b=\sqrt{\frac{3}{2}\frac{D}{G}}\sim \tau^{\nu/2}$; $d=\sqrt{\frac{1}{2}\frac{D}{G}}\sim\tau^{\nu/2}$.

\begin{figure}[!t]
	\centerline{
\includegraphics[angle=0,width=0.53\textwidth]{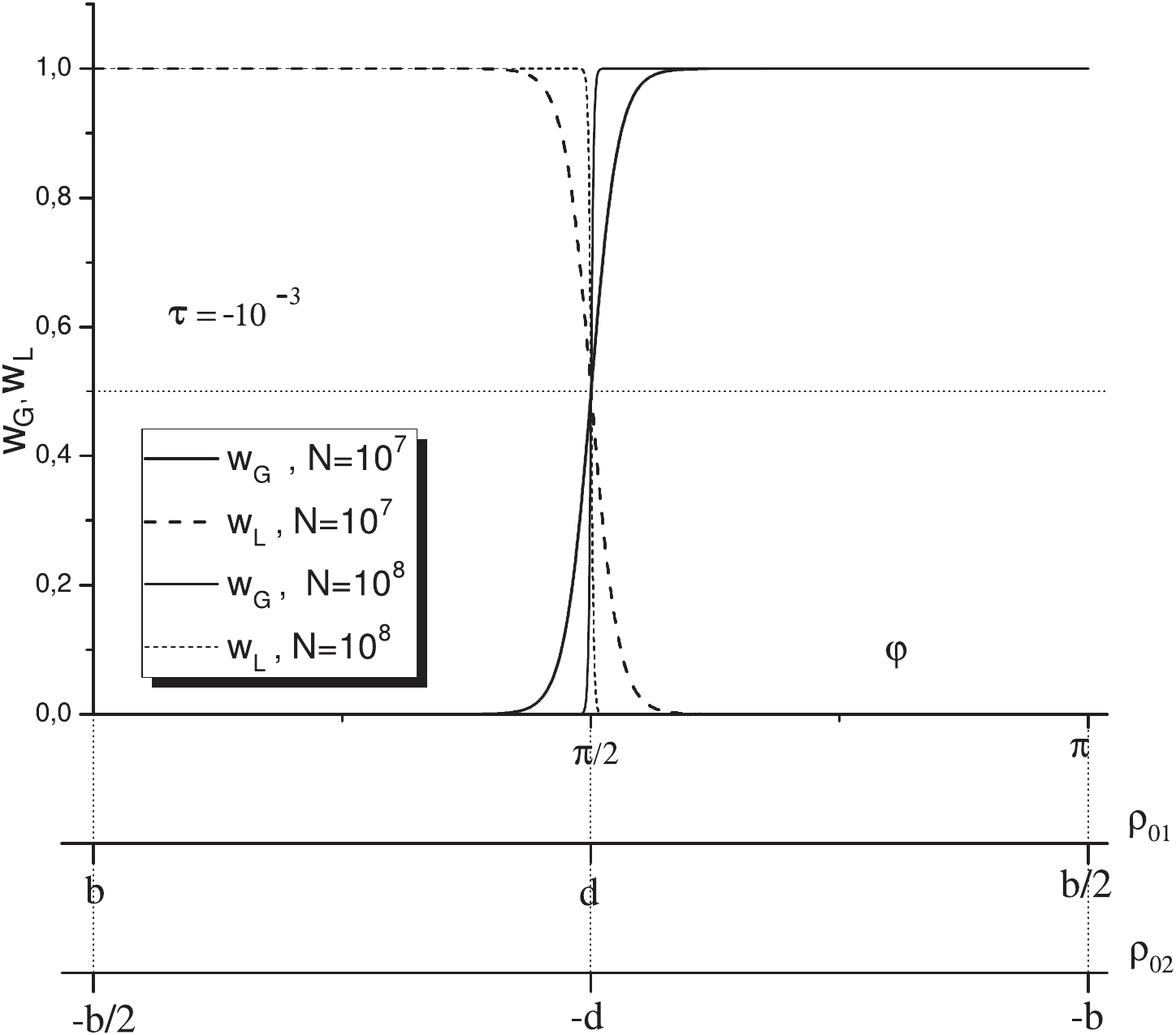}
}
	\caption{Probabilities $w_{\textrm{l}}$ and $w_{\textrm{g}}$ for single-phase systems,
of liquid ($w_{\textrm{l}}$) and of gas ($w_{\textrm{g}}$), respectively, as functions
of the angle $\varphi$. Two bottom axes are to represent values $\rho_{01}(\varphi)$
and $\rho_{02}(\varphi)$.}  \label{fig:Prob}
\end{figure}

\begin{figure}[!b]
	\centerline{
\includegraphics[angle=0,width=0.65\textwidth]{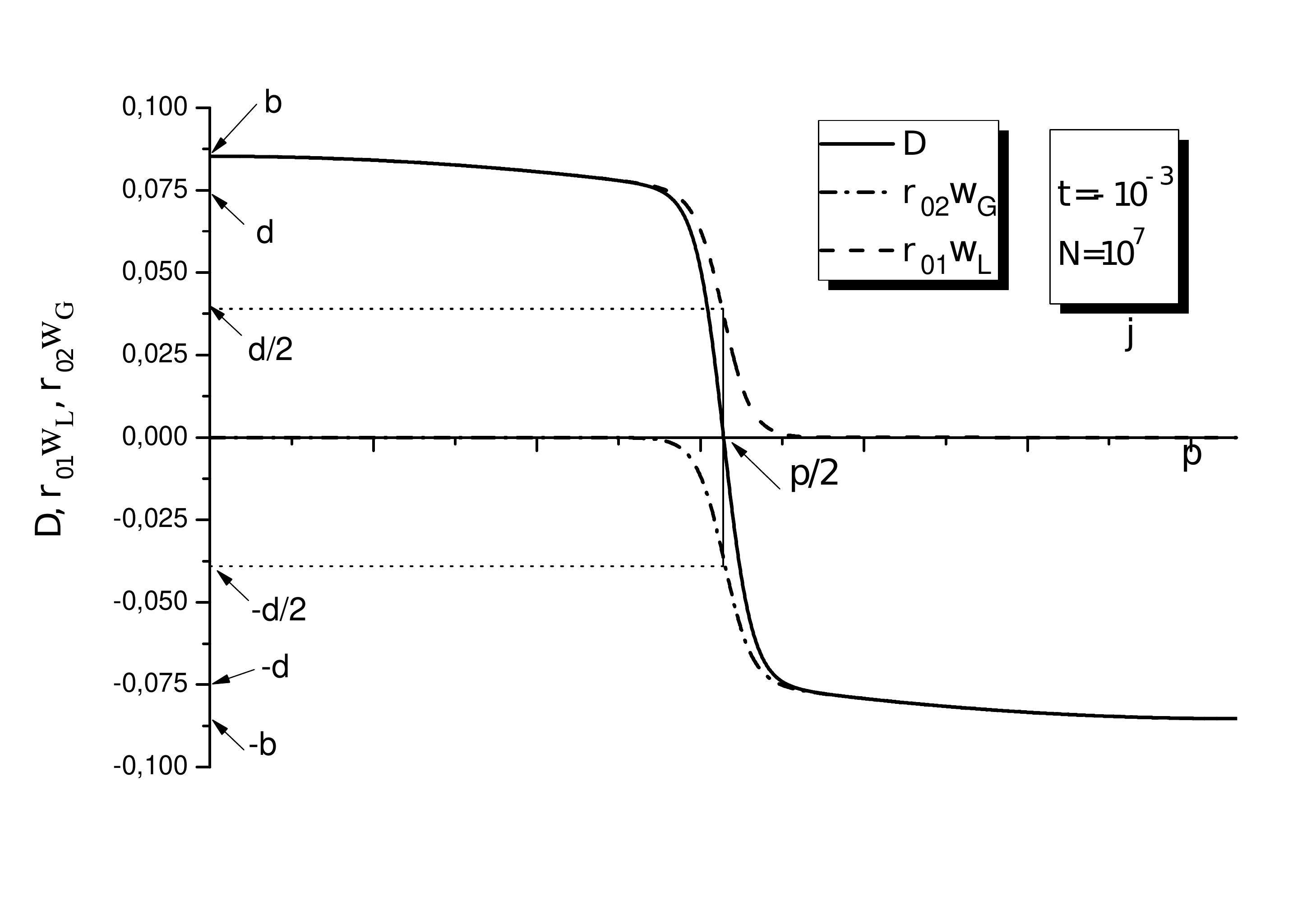}
}
	\caption{Plot for $\Delta=\Delta_\textrm{g}+\Delta_\textrm{l}$ and $\Delta_\textrm{g}=w_\textrm{g}\rho_{02}$;
$\Delta_\textrm{l}=w_\textrm{l}\rho_{01}$; $\rho_{02}=b\cos{\frac{\varphi+2\pi}{3}}$; $\rho_{01}
=b\cos{\frac{\varphi}{3}}$; $0\leqslant \varphi\leqslant \pi$. The vertical fragment $d/2\div-d/2$
joins the point of the largest densities of gas with that of the smallest density of liquid in the two-phase system.
	} \label{fig:Delta}
\end{figure}

In formulas~(\ref{eq3.11}),~(\ref{eq3.13a}) and~(\ref{eq3.13b})
 we give important points for functions $w_{\textrm{g}}$, $w_{\textrm{l}}$ and $\Delta_\textrm{g}$,
 $\Delta_\textrm{l}$. In fact, we should be driven by expressions~(\ref{eq3.10}) or even by more precise
 solutions of equation~(\ref{eq3.3}). In figure~\ref{fig:Prob}, the graphics of $w_{\textrm{g}}$
 and $w_{\textrm{l}}$ are presented as functions of the angle $\varphi$.

Attention should be paid to the fact that the function $\Delta$ as a sum
$\Delta = w_{\textrm{g}} \rho_{02} + w_{\textrm{l}} \rho_{01} = \Delta_{\textrm{g}}
+ \Delta_{\textrm{l}}$ turns into zero at $\Delta_{\textrm{g}} = - d/2$ and
$\Delta_{\textrm{l}} = d/2$. In figure~\ref{fig:Delta}, the line $(-d/2 \div d/2)$ is interpreted as the line of the gas-liquid,
first-order phase transition. The points $\Delta_\textrm{g}|_{\varphi=\pi/2}=-d/2$
and  $\Delta_\textrm{l}|_{\varphi=\pi/2}=d/2$ do not lie on the curve
$\Delta(\varphi)|_{\varphi=\pi/2}$. The sum of their values corresponds
to the point $\Delta(\varphi)=0$ for $\varphi=\frac{\pi}{2}$.
The expression $\Delta_\textrm{g}=-d/2$ consists of two factors: $\rho_{02}=-d$ is the
end of the principal maxima of the function $\exp [NE(\rho_0)]$ located upon the
gaseous branch $\rho_{02}(\varphi)$, and $w_\textrm{g}=1/2$ that characterises the
probability of arising of the value $\rho_{02}$. For the point $\Delta_\textrm{l}=d/2$,
it corresponds to $w_\textrm{l}=1/2$ and the value $\rho_{01}=d$, which starts the curve
of the principal maxima of $\exp\left[NE(\rho_0)\right]$ located upon the liquid
branch $\rho_{01}(\varphi)$. Note, that the density of either gaseous or liquid phase
is independent of either  $\rho_{01}$ or $\rho_{02}$, but depends on $\Delta$,
according to~(\ref{eq3.13})--(\ref{eq3.13b}). A jump of the density at the phase
transition is connected with the points $\rho_{01}=d$ and $\rho_{02}=-d$, given in
figure~\ref{fig:mu_w} for $\mu^{*}=\mu^{*}(\rho_0)$, but its magnitude depends on
the values of products $w_\textrm{g}\rho_{02}$ and $w_\textrm{l}\rho_{01}$ in these points,
in other words, on the probabilities $w_\textrm{g}(-d)=\frac{1}{2}$ and $w_\textrm{l}(-d)=\frac{1}{2}$
as well. Thus, we have
\be
\label{eq3.14}
\eta_\textrm{g} |_{\varphi={\pi}/{2}}=\eta_\textrm{c}(-d/2)+\eta_\textrm{c} \qquad {\text{and}} \qquad
\eta_\textrm{l} |_{\varphi={\pi}/{2}}=\eta_\textrm{c}(d/2)+\eta_\textrm{c}
\ee
and the jump of density equals $\eta_\textrm{c} d$.

\begin{figure}[!t]
	\centerline{
\includegraphics[angle=0,width=0.7\textwidth]{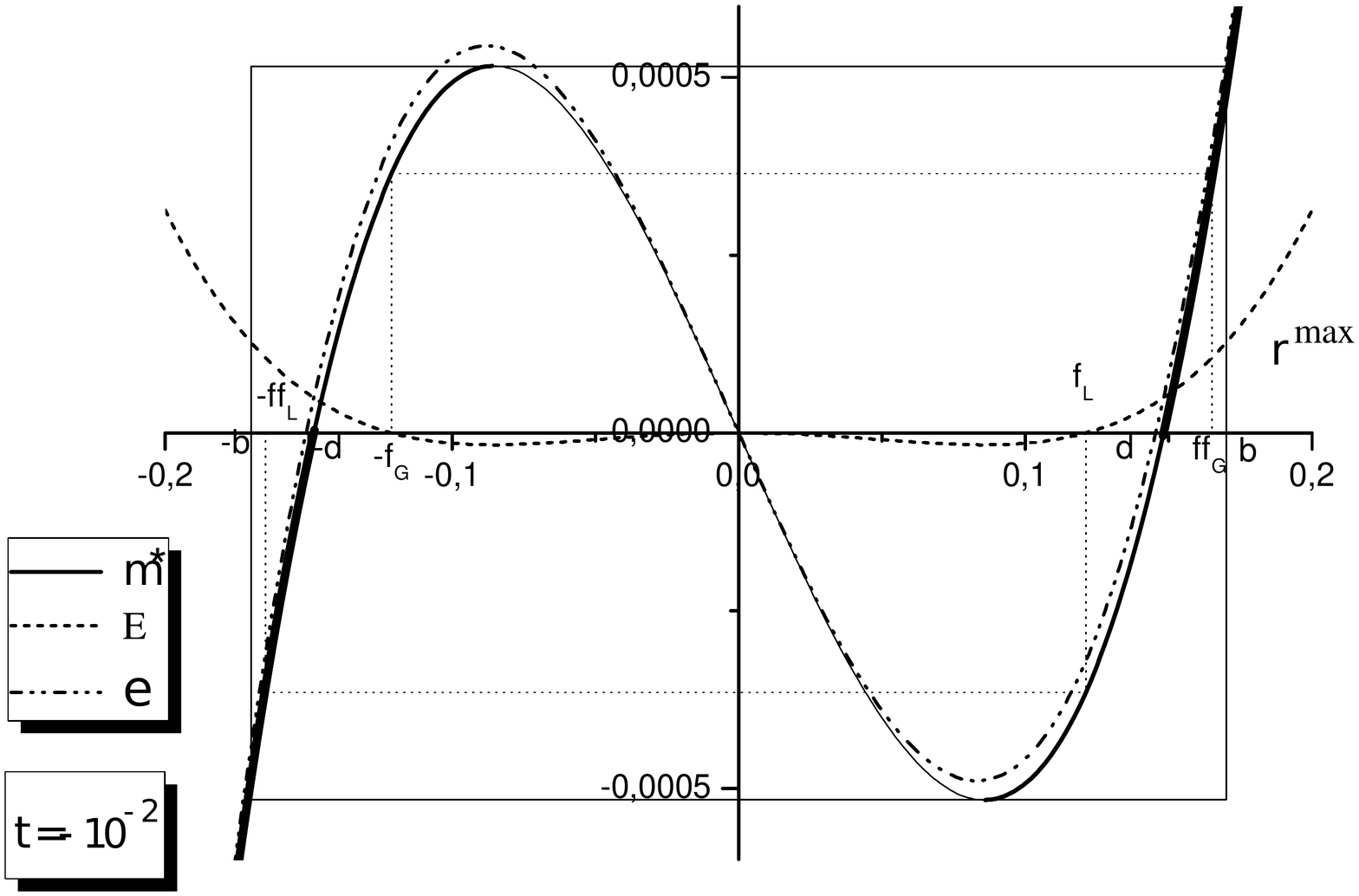}
}
	\caption{The quantities ${\cal E}$, $E$ and $\mu^*$ as functions of
$\rho_{\rm max}=b\cos{\frac{\varphi+2\pi n}{3}}, n=1,2,3$ within the
region $Q<0$ [see equations~(\ref{eq23}),~(\ref{eq27}) and~(\ref{eq28})].
Thick lines represent the principal maxima for $E(\rho)$. Thin lines represent
the secondary maxima. For gaseous branch $n=1$, the points $-b, -d, -b/2$
correspond to values $\varphi=0, \pi/2, \pi$; for liquid branch $n=0$,
the points $b, d$ and $b/2$ correspond to the values $\varphi=0, \pi/2, \pi$, respectively.} \label{fig:three_quan}
\end{figure}

Also note that $\ddot{E}(\rho_0)$, which enter the coefficients
$c_{\textrm{g}}$ and $c_{\textrm{l}}$ in~(\ref{eq3.10}), are even functions
of $\rho_0$, $\ddot{E}(\rho_0) = 2D - 12G \rho_0^2$. Thus, $\ddot{E}(b)
= \ddot{E}(-b)$, $\ddot{E}(d) = \ddot{E}(-d)$. In general,
for the points $\rho_{02}$ and $\rho_{01}$, which are symmetric relative to the rectilinear
diameter $\Delta = 0$, there is $\ddot{E}(\rho_{02})/\ddot{E}(\rho_{01}) = 1$.

In figure~\ref{fig:three_quan}, there are presented the curves ${\cal E}=\ln{\Xi_{\rho_0}}$,
$E(\rho_0)$ and $\mu^{*}(\rho_0)$ as functions of $\rho_{\rm max}=b\cos{\frac{\varphi+2\pi n}{3}}$
for $n=0,1$, $\rho_{\rm max}$ correspond to maximum values of the function $E(\rho)$ in
equation~(\ref{eq3.1}) for $\ln{\Xi_{\rho_0}}$.

The very thin line represents the minima of the function $E(\rho)$, which is the region of
thermodynamic instability. The quantity $\exp [NE(\rho)]$ characterises the probability
measure for a state to exist. The points $ff_\textrm{l}$ and $ff_\textrm{g}$, which are the ends of
$f_{\textrm{l}} \div ff_{\textrm{l}}$ and $f_{\textrm{g}} \div ff_{\textrm{g}}$ determine
the limits for stable density fluctuations of liquid type arising in the gaseous phase
and of gaseous type in a liquid. The points $-{b}/{2}$ and ${b}/{2}$ determine
the limits of secondary maxima for gaseous and liquid phases, respectively.

The points $-d/2$ and $d/2$ are the points of fixed densities for gas in the gaseous
phase and for liquid in the liquid stage in the two-phase system, which arise at the
first-order phase transition. In figure~\ref{fig:three_quan}, the closeness of the
curves  $\mu^*(\rho_{\rm max})$ and ${\cal E}(\rho_{\rm max})$ is obvious, which
makes it possible, in what follows, to write  ${\cal E}(\Delta) \approx \mu^*(\Delta)$.

Let us return to the initial expression for $\Xi_{\rho_0}$, given in~(\ref{eq3.1})
for $Q < 0$. Similarly to the case of $Q > 0$, in equations~(\ref{eq1.8})--(\ref{eq1.12})
we write $\ln \Xi_{\rho_0}$ in the form:
\be
\label{eq3.15}
\ln \Xi_{\rho_0} = N {\cal E}(\Delta),
\ee
keeping the notation of equations~(\ref{eq1.8})--(\ref{eq1.12}). However, now
expression~(\ref{eq3.10}) is taken for $\Delta$.

The quantity ${\cal E}(\Delta)$, where $\Delta$ is a solution to equation~(\ref{eq3.2})
specified in~(\ref{eq3.10}), is a curve that envelops the principal and the secondary maxima of the function $E(\rho)$.

It follows from figure~\ref{fig:three_quan} that the curves ${\cal E}(\rho_{\rm max})$
and $\mu^*(\rho_{\rm max})$ stretch along each other. This is not surprising because
${\cal E}(\rho_{\rm max})$ consists of the sum $\mu^*(\rho_{\rm max})$
and expressions $D\rho_{\rm max}^2 - G\rho_{\rm max}^4$. In the critical
 region $\mu^* \sim \tau^{5/2\nu}$, and $D\rho_{\rm max}^2 - G\rho_{\rm max}^4 \sim \tau^{3\nu}$.

Therefore, in what follows, we shall write
\bea
\label{eq3.16}
&& {\cal E}\phantom{^*}(\Delta) = \mu^*(\Delta) + D\Delta^2 - G\Delta^4, \nonumber\\
&& \mu^*(\Delta) = -2D\Delta + 4G\Delta^3,\nonumber \\
&& {\cal E}\phantom{^*}(\Delta) \simeq \mu^*(\Delta)
\eea
and for the equation of state, taking into account~(\ref{eq3.15}) and~(\ref{eq3.16}):
\bea
\label{eq3.17}
&& P_{\rho_0} = \frac{\Theta}{V} \ln \Xi_{\rho_0}\,, \nonumber\\
&& P_{\rho_0} = \Theta \frac{N}{V} \mu^*(\Delta), \qquad -b \leqslant \Delta \leqslant b .
\eea
The value of $\Delta$ is given in~(\ref{eq3.10}). For $\mu^{*}$ we may
take two equivalent expressions. In the first one, according to~(\ref{eq27})
\[
\mu^{*} = a \cos{\varphi}, \qquad  0 \leqslant \varphi \leqslant \pi,
\]
for the second one according to~(\ref{eq20})
\[
\mu^{*} = -2D\rho_{\rm max} +4G\rho_{\rm max}^3\,,
\]
where, based on~(\ref{eq27}), $\rho_{\rm max}$ may take on two values:
\[
\rho_{\rm max} = \rho_{02} = b\cos{\frac{\varphi+2\pi}{3}}, \qquad 0\leqslant \varphi \leqslant \pi,
\]
and this means the branch of the gaseous type, and
\[
\rho_{\rm max} = \rho_{01} = b\cos{\frac{\varphi}{3}}\,, \qquad 0\leqslant \varphi \leqslant \pi,
\]
describes the branch of the liquid type.

\section{Properties of functions ${\cal{E}} (\Delta)$ and $\mu^{*}(\Delta)$ in the region of $Q<0$}
\label{sec:func_properties}
In the region of $Q<0$ for $0\leqslant \varphi \leqslant \pi$, there are two
solutions to equations~(\ref{eq20}),~(\ref{eq21}): the solution of
the gaseous type $\rho_{02}$ and that of the liquid type $\rho_{01}$.
Moreover, when $\rho_{02}(\varphi)$ for $\pi/2\leqslant\varphi\leqslant\pi$
describes the principal maximum, in the same interval of values $\varphi$
the solution $\rho_{01}(\varphi)$ describes the secondary maximum. Inverse
situation is seen in the region where $0\leqslant \varphi \leqslant \pi/2$: the principal
maximum is due to $\rho_{01}$, and the secondary one is due to $\rho_{02}$.
In general, this is governed by the probabilities  $w_\textrm{g}$ and $w_\textrm{l}$ given
in~(\ref{eq3.10}). In order to take into account both maxima, in the
expression for $\mu^{*}(\rho_{\rm max})$ we should write:
\be
\label{eq4.1}
\mu^{*} =\mu^{*}(\Delta(\varphi)) = -2D\Delta + 4G\Delta^3,
\ee
where
\bea
&&\Delta = \Delta_\textrm{g} +\Delta_\textrm{l}\,, \nonumber\\
&&\Delta_\textrm{g} = w_\textrm{g}\rho_{02}\,, \qquad \Delta_\textrm{l} = w_\textrm{l}\rho_{01}\,.\nonumber
\eea

\begin{figure}[!t]
	\centerline{
\includegraphics[angle=0,width=0.65\textwidth]{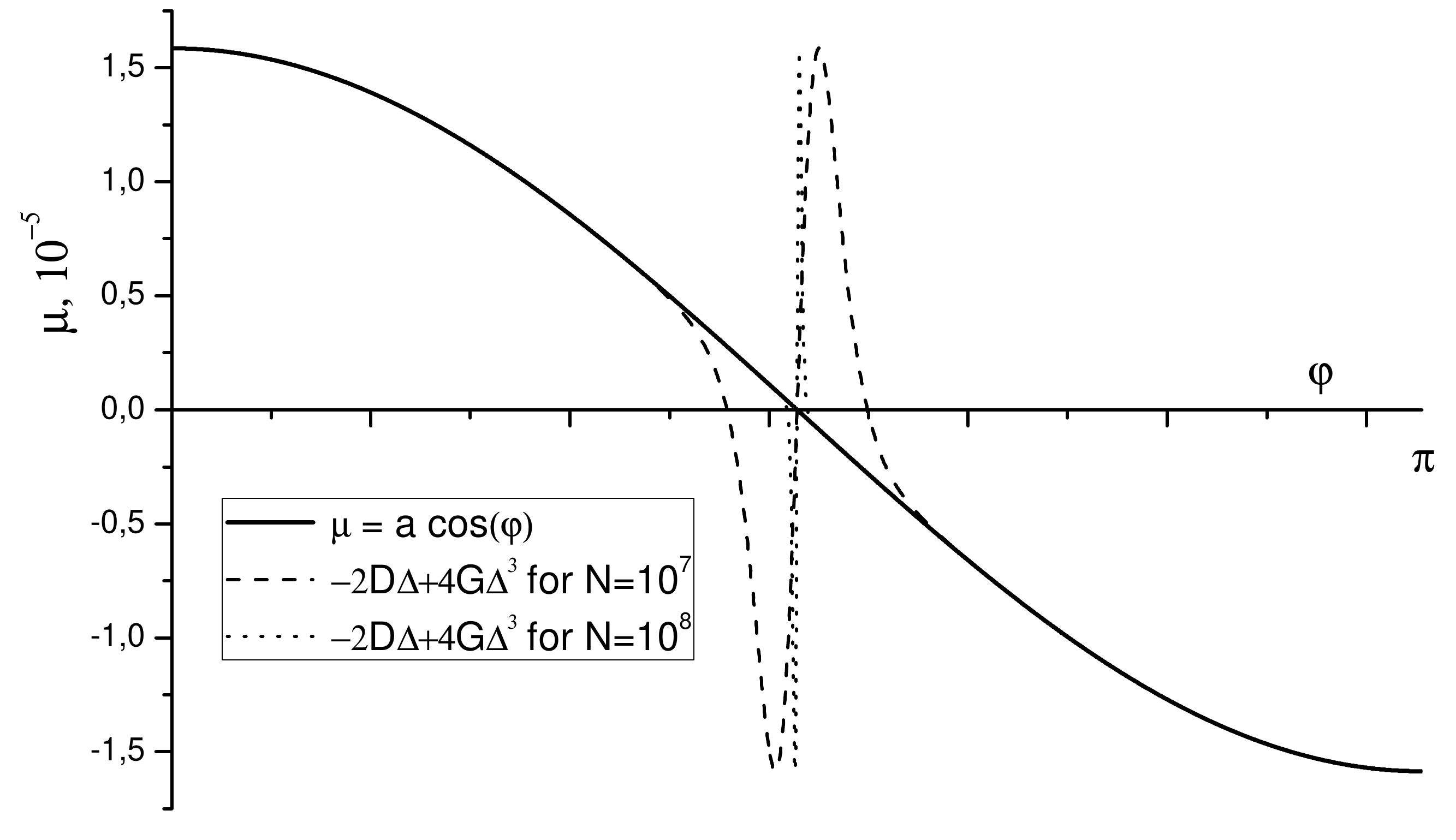}
}
	\caption{The comparison of the curves $\mu^{*}=a\cos{\varphi}$ and
$\mu^{*}=-2D\Delta(\varphi)+4G\Delta(\varphi)^3$ as functions of the
angle $\varphi$, $0\leqslant\varphi\leqslant\pi$; $N=10^7$ and $N=10^8$. As $N$
grows, the attitudes of maximum and minimum remain unchanged, the
widths go to zero.  See also figures~\ref{fig:Prob} and~\ref{fig:Delta}
for $w_\textrm{g}$ and $w_\textrm{l}$ as well as for $\Delta_\textrm{g}$, $\Delta_\textrm{l}$ and
$\Delta=\Delta_\textrm{g}+\Delta_\textrm{l}$. } \label{fig:mu_param}
\end{figure}

Figure~\ref{fig:mu_param} presents the curves for the generalized chemical
potential $\mu^{*}(\varphi)$ as functions of $\Delta(\varphi)$:
one curve is for  $\mu^{*}=a\cos{\varphi}$ and the other is for $\mu^{*} = -2D\Delta(\varphi)
+4G\Delta(\varphi)^3$. We are going to work with the latter one. They almost fully
coincide except for the narrow interval of values $\varphi$ near $\varphi=\pi/2$.
It follows from figures~\ref{fig:mu_w} and~\ref{fig:mu_param}, in the region of $Q<0$,
that in the process of isothermal compression of the gas along $\rho_{02}(\varphi)$
for $\pi/2\leqslant\varphi\leqslant \pi$, the system via $\Delta_\textrm{l}$ enters the region of
liquid-like values of $\mu^{*}(\Delta_\textrm{g}+\Delta_\textrm{l})$. In other words, at
compression of the gaseous phase in the region $Q<0$, the density fluctuations of
the liquid type emerge. Ultimately, at $\varphi=\pi/2$, a droplet of liquid appears in the gaseous phase.

What to do next depends on the type of the shell the droplet possesses when
the two-phase system arises. The simplest case is when the shell of the droplet
is presented as a geometric sphere confining the droplet. The shell does not
contain particles. The question arises whether this is physically justified. In fact,
the shell of the droplet has a volume, and is of the form of a spherical layer with a
non-neglected thickness. All of them, i.e., two phases and the surface layer, should be
present in the right-hand side of~(\ref{eq4.1}). The left-hand side of~(\ref{eq4.1}),
which initially is of the form ${PV}/{\Theta} = \ln{\Xi_{\rho_0}}$, is written as
a sum of terms dependent on the right-hand side of~(\ref{eq4.1}), in the expression $\ln{\Xi_{\rho_0}}$.

Returning to~(\ref{eq4.1}), for $\Delta(\varphi)$, in accordance with~(\ref{eq3.10}), we write
\[
\Delta = \Delta_\textrm{g} + \Delta_\textrm{l}\,, \qquad \Delta_\textrm{g}=w_\textrm{g}\rho_{02}\,, \qquad \Delta_\textrm{l}=w_\textrm{l}\rho_{01}\,.
\]
Then,
\be
\label{eq4.2}
{\cal{E}}(\Delta) = {\cal{E}}(\Delta_\textrm{g}+\Delta_\textrm{l}) =
\mu^{*}(\Delta_\textrm{g}+\Delta_\textrm{l}) + D(\Delta_\textrm{g}+\Delta_\textrm{l})^2 -G(\Delta_\textrm{g}+\Delta_\textrm{l})^4.
\ee
Let us write this as a sum
\bea
\label{eq4.3}
&&{\cal{E}}\phantom{^{*}}(\Delta_\textrm{g}+\Delta_\textrm{l})={\cal{E}}_\textrm{g}(\Delta_\textrm{g},\Delta_\textrm{l})+
{\cal{E}}_\textrm{l}(\Delta_\textrm{g},\Delta_\textrm{l}), \nonumber \\
&&\mu^{*}(\Delta_\textrm{g}+\Delta_\textrm{l})=\mu^{*}_\textrm{g}(\Delta_\textrm{g},\Delta_\textrm{l})+
\mu^{*}_\textrm{l}(\Delta_\textrm{g},\Delta_\textrm{l}).
\eea
By doing so, we make no changes compared to either~(\ref{eq4.1}) or~(\ref{eq4.2}).

The curves for $\Delta_\textrm{g}$ and $\Delta_\textrm{l}$ are presented in figure~\ref{fig:Prob}.
Take the point $\Delta_\textrm{g}=-d/2$ on the curve $\Delta_\textrm{g}$, and the point $\Delta_\textrm{l}=d/2$
on the curve $\Delta_\textrm{l}$. Both correspond to equal probabilities $w_\textrm{g}=w_\textrm{l}=1/2$ for $w_\textrm{g}$
and $w_\textrm{l}$ in figure~\ref{fig:Prob}. Obviously,
\be
\label{eq4.4}
\Delta_\textrm{g}(-d/2)+\Delta_\textrm{l}(d/2) = \Delta(\pi/2)=0.
\ee
In figures~\ref{fig:mu_w},~\ref{fig:three_quan},~\ref{fig:mu_param} the points
$\Delta(\pi/2)=0$ and $\mu^{*}(\Delta(\varphi))_{\Delta=0}=0$ correspond to the
transition from the gaseous part $\mu^{*}(\Delta(\varphi))$ to the liquid part
$\mu^{*}(\Delta(\varphi))$. At this point,
\be
\label{eq4.5}
\Delta=0, \qquad \mu^{*}(\Delta)=0, \qquad {\cal{E}}(\Delta)=0.
\ee

Our aim is to show that the liquid-gas phase transition occurs along the line
$\Delta_\textrm{g}=-d/2 \div \Delta_\textrm{l}=d/2$, and that condition~(\ref{eq4.5}) is obeyed
starting at $\Delta_\textrm{g}=-d/2$ and finishing at $\Delta_\textrm{l}=d/2$, and that this
process can be described within the Gibbs statistics. Since the first-order phase
transition at a constant temperature $\tau$ occurs due to the external latent work
of the pressure, we should demonstrate that at $\varphi=\pi/2$, and $w_\textrm{l}=w_\textrm{g}=1/2$, the following is true
\be
\mu_\textrm{g}^*(\varphi)|_{\varphi=\pi/2} = \mu_\textrm{g}^*(\Delta_\textrm{g},\Delta_\textrm{l}) 
\Big|_{
\begin{smallmatrix}\Delta_\textrm{g}=-d/2 \\ \Delta_\textrm{l}=d/2 \hspace{2.2mm}
\end{smallmatrix}}=0; \nonumber
\ee
\be
\label{eq4.6}
\mu_\textrm{l}^*(\varphi)|_{\varphi=\pi/2} = \mu_\textrm{l}^*(\Delta_\textrm{g},\Delta_\textrm{l}) 
\Big|_{
\begin{smallmatrix}\Delta_\textrm{g}=-d/2 \\ \Delta_\textrm{l}=d/2\hspace{2.2mm}
\end{smallmatrix}}=0; \nonumber
\ee
\be
{\cal{E}}_\textrm{g}(\varphi)|_{\varphi=\pi/2} = {\cal{E}}_\textrm{g}(\Delta_\textrm{g},\Delta_\textrm{l}) 
\Big|_{
\begin{smallmatrix}\Delta_\textrm{g}=-d/2 \\ \Delta_\textrm{l}=d/2\hspace{2.2mm}
\end{smallmatrix}}=0; \nonumber
\ee
\be
{\cal{E}}_\textrm{l}(\varphi)|_{\varphi=\pi/2} = {\cal{E}}_\textrm{l}(\Delta_\textrm{g},\Delta_\textrm{l}) 
\Big|_{
\begin{smallmatrix}\Delta_\textrm{g}=-d/2 \\ \Delta_\textrm{l}=d/2\hspace{2.2mm}
\end{smallmatrix}}=0.
\ee

For this reason, we return to~(\ref{eq4.2}) and~(\ref{eq4.3}). Let us show that
equations~(\ref{eq4.5}),~(\ref{eq4.6}) do take place along the
line $\mu^{*}(\Delta_\textrm{g},\Delta_\textrm{l})=0$ for $\Delta$ in the interval
$\Delta_\textrm{g}=-d/2 \div \Delta_\textrm{l}=d/2$. We write the expressions
$\mu^{*}(\Delta_\textrm{g}+\Delta_\textrm{l})$ and ${\cal{E}}(\Delta_\textrm{g}+\Delta_\textrm{l})$,
selecting the mixed products $\Delta_\textrm{g}^m\Delta_\textrm{l}^n$ and separating
them for $\mu_\textrm{g}$ and $\mu_\textrm{l}$ and for $ {\cal{E}}_\textrm{g} $ and $ {\cal{E}}_\textrm{l}$
\begin{align}
\label{eq4.7}
\mu_\textrm{g}^*(\Delta_\textrm{g},\Delta_\textrm{l}) &= -2D\Delta_\textrm{g}+4G\Delta_\textrm{g}^3+12G\Delta_\textrm{g}\Delta_\textrm{l}^2\,, \nonumber \\
\mu_\textrm{l}^*(\Delta_\textrm{g},\Delta_\textrm{l}) &= -2D\Delta_\textrm{l}+4G\Delta_\textrm{l}^3+12G\Delta_\textrm{l}\Delta_\textrm{g}^2\,.
\end{align}
Obviously, the sum of these expressions is equal to the initial form of $\mu^{*}(\Delta_\textrm{g}+\Delta_\textrm{l})$ in~(\ref{eq4.2}). Next,
\bea
\label{eq4.8}
{\cal{E}}_\textrm{g}(\Delta_\textrm{g},\Delta_\textrm{l}) &=& \mu_\textrm{g}^*(\Delta_\textrm{g},\Delta_\textrm{l}) +D\Delta_\textrm{g}^2-G\Delta_\textrm{g}^4+D\Delta_\textrm{g}\Delta_\textrm{l} \nonumber\\
&&-\,G\left[2\left(\Delta_\textrm{g}^3\Delta_\textrm{l}+\Delta_\textrm{g}\Delta_\textrm{l}^3\right) +3\Delta_\textrm{g}^2\Delta_\textrm{l}^2\right], \nonumber\\
{\cal{E}}_\textrm{l}(\Delta_\textrm{g},\Delta_\textrm{l}) &=& \mu_\textrm{l}^*(\Delta_\textrm{g},\Delta_\textrm{l}) +D\Delta_\textrm{l}^2-G\Delta_\textrm{l}^4+D\Delta_\textrm{g}\Delta_\textrm{l} \nonumber\\
&&-\,G\left[2\left(\Delta_\textrm{g}^3\Delta_\textrm{l}+\Delta_\textrm{g}\Delta_\textrm{l}^3\right) +3\Delta_\textrm{g}^2\Delta_\textrm{l}^2\right].
\eea
The sum of these expressions equals $ {\cal{E}}_\textrm{g}(\Delta_\textrm{g},\Delta_\textrm{l}) $ in~(\ref{eq4.2}).
By substitution $ \Delta_\textrm{g}=w_\textrm{g}\rho_{02} $ and $ \Delta_\textrm{l}=w_\textrm{l}\rho_{01} $ we make sure
that on both ends of the line of phase transition
\be
\label{eq4.9}
\Delta_\textrm{g}=-d/2 \div \Delta_\textrm{l}=d/2, \qquad \varphi = \pi/2,
\ee
(see figures~\ref{fig:Prob} and~\ref{fig:Delta}) equalities~(\ref{eq4.6}) hold:
\begin{align}
\label{eq4.10}
\mu^{*}_\textrm{g} (-d/2,d/2)&=0, &\mu^{*}_\textrm{l} (-d/2,d/2)&=0, \nonumber \\
{\cal{E}}_\textrm{g} (-d/2,d/2)&=0, & {\cal{E}}_\textrm{l} (-d/2,d/2)&=0.
\end{align}

We conclude that the line $-d/2\div d/2$ has the properties of the line of the
first-order phase transition in the gas-liquid system, at arbitrarily large $N$.

The expressions for generalized chemical potentials $\mu^{*}_\textrm{g}$ and $\mu^{*}_\textrm{l}$
as well as for ${\cal{E}}_\textrm{g}$ and ${\cal{E}}_\textrm{l}$ consist of both ``pure'' phase terms
such as $-D\Delta_\textrm{g}+4G\Delta_\textrm{g}^3$, and ``mixed'' products such as $ \Delta_\textrm{g}\Delta_\textrm{l}^2$,
$ \Delta_\textrm{g}^2\Delta_\textrm{l}$, $ \Delta_\textrm{g}\Delta_\textrm{l} $ etc. On the line $-d/2\div d/2$, the ``mixed''
products are of the same order of magnitude  as the ``pure'' ones. Note that in~(\ref{eq4.8}),
quantities $ \mu_\textrm{g}^* $, $ \mu_\textrm{l}^* $ are proportional to $\tau^{5\nu/2}$, and the expression
$D\Delta^2-G\Delta^4$ is proportional to $\tau^{3\nu}$.

\begin{figure}[!t]
	\centerline{
\includegraphics[angle=0,width=0.65\textwidth]{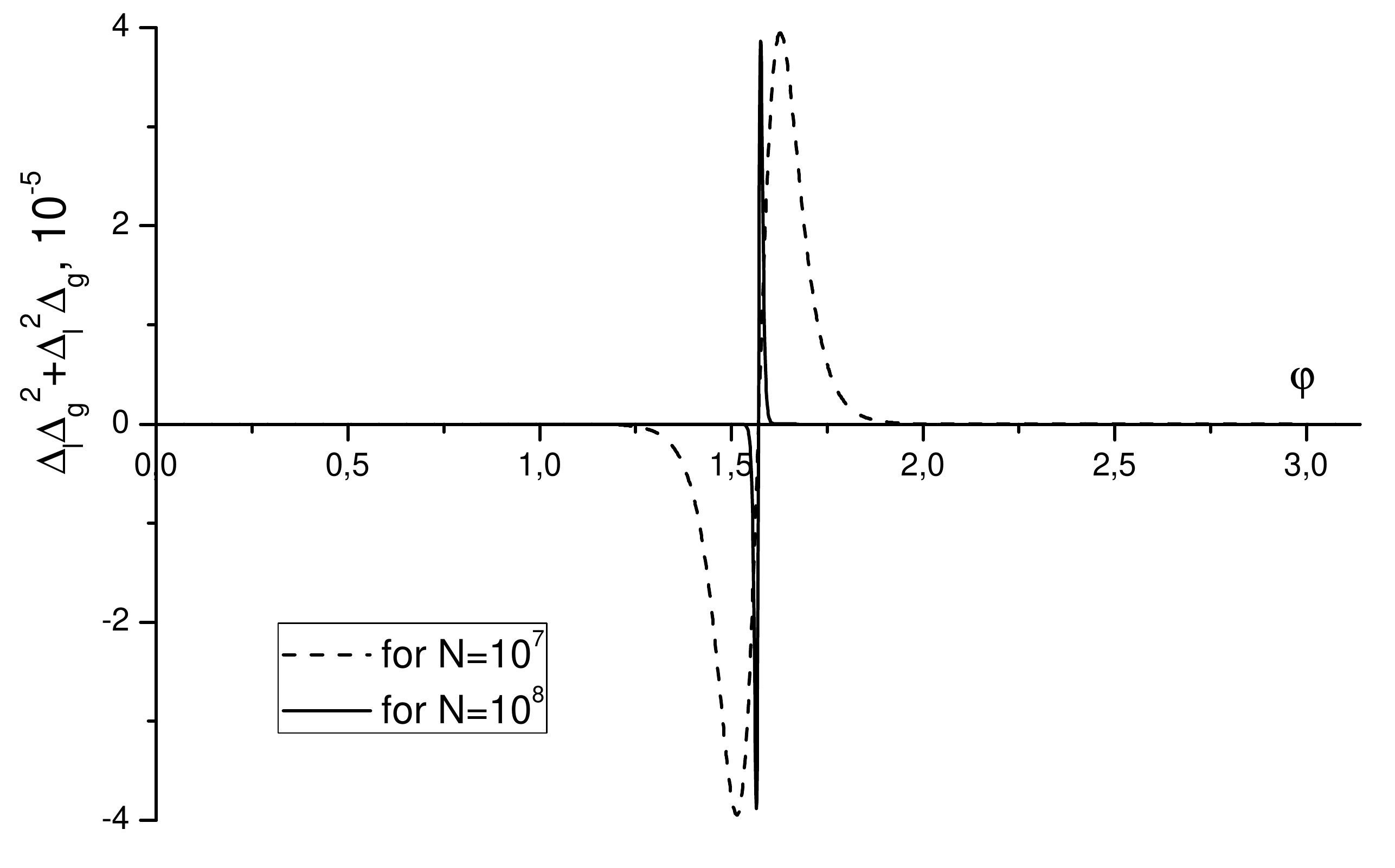}
}
	\caption{The region of the surface layer described by the products
		$ \Delta_\textrm{g}\Delta_\textrm{l}^2$, $ \Delta_\textrm{g}^2\Delta_\textrm{l}$ in the expressions
for ${\cal{E}}(\Delta_\textrm{g},\Delta_\textrm{l})$ and $\mu^{*}(\Delta_\textrm{g},\Delta_\textrm{l})$
in~(\ref{eq4.7}),~(\ref{eq4.8}). The width of this region of values $\Delta_\textrm{g}^m\Delta_\textrm{l}^n$
depends on $N$. At  $N$ increasing, this region narrows and joins the surface $\varphi=\pi/2,\, \mu^{*}=0$.
In addition, the magnitudes of maxima remain constant, and the effect of the mixed products
on the requirement~(\ref{eq4.5}),~(\ref{eq4.6}) does not change.
	} \label{fig:7a}
\end{figure}

In figure~\ref{fig:7a}, there are presented the curves for the mixed products
$ \Delta_\textrm{g}\Delta_\textrm{l}^2$, $ \Delta_\textrm{g}^2\Delta_\textrm{l}$. Their non-zero values
are located near the limiting value $\varphi=\pi/2$ and characterise the
width of transition layer between gas and liquid. The size of the region of
values  $\Delta_\textrm{g}^m\Delta_\textrm{l}^n$ depends on $N$. At $N$ increasing, it
narrows and the region of the surface shell joins the surface $\mu^{*}=0$,
for the angle $\varphi=\pi/2$. Hence, the magnitude of the maxima keep their
values unchanged. As a result, the effect of the mixed products on the requirements~(\ref{eq4.5}),~(\ref{eq4.6}) remains unchanged.

\begin{figure}[!b]
	\centerline{
\includegraphics[angle=0,width=0.6\textwidth]{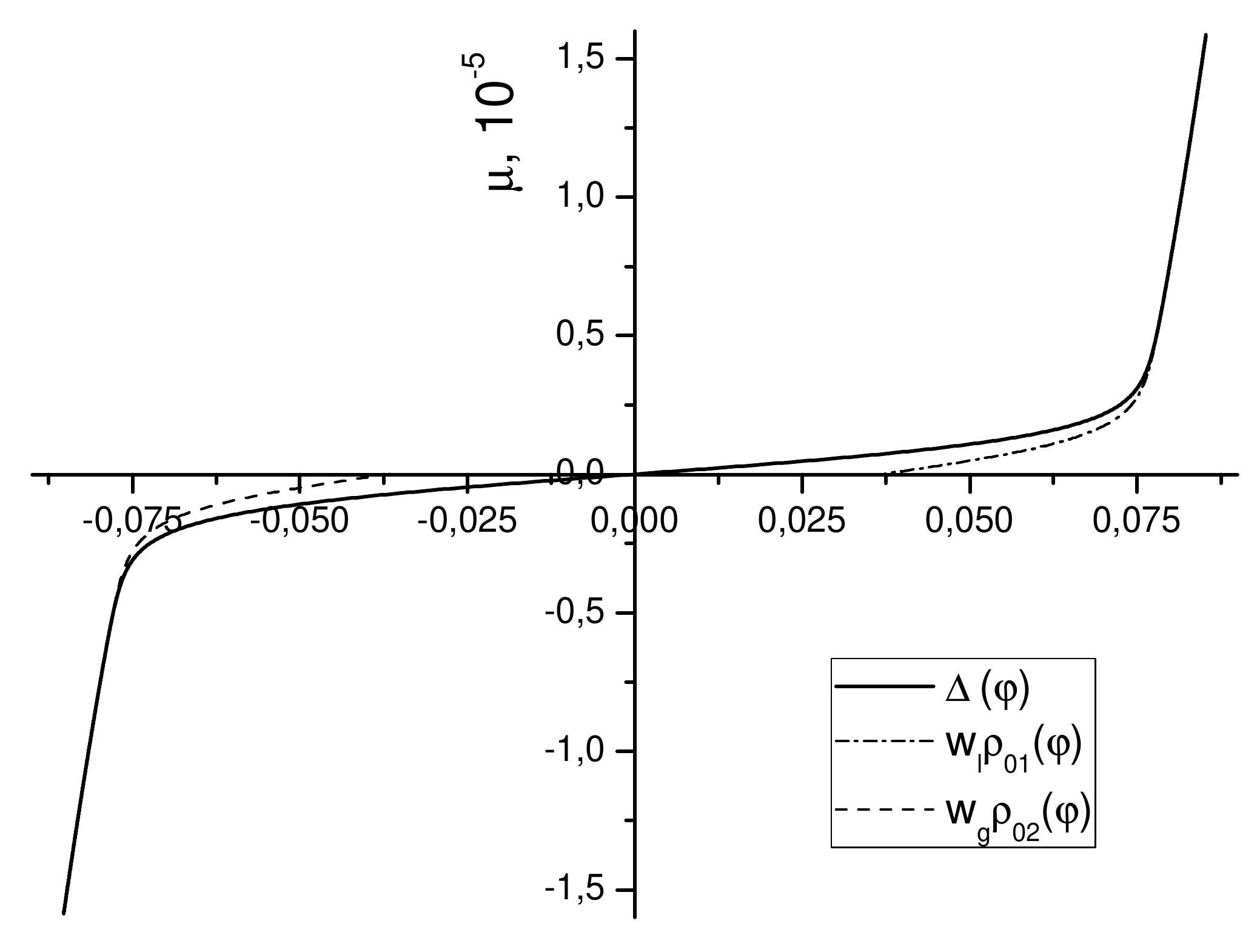}
}
	\caption{The curve of $\mu^{*}(\Delta)$ as a relult of combination of
$\mu^{*}=a\cos{\varphi}$ and $ \Delta = w_\textrm{g}\rho_{02}(\varphi) + w_\textrm{l}\rho_{01}(\varphi)$.
The influence of the mixed products $12G\Delta_\textrm{g}\Delta_\textrm{l}^2$ on nullifying the branch
$ \mu^*_\textrm{g} (\Delta_\textrm{g},\Delta_\textrm{l}) $ in the point $\Delta_\textrm{g}=-d/2$ and of
$12G\Delta_\textrm{g}\Delta_\textrm{l}^2$ on nullifying the branch $ \mu^*_\textrm{l} (\Delta_\textrm{g},\Delta_\textrm{l}) $
in the point $\Delta_\textrm{l}=d/2$ is shown.
	} \label{fig:8}
\end{figure}

It is seen in equation~(\ref{eq4.7}) that the mixed product $12G\Delta_\textrm{g}\Delta_\textrm{l}^2$
forces the gaseous branch $ \mu^{*}_\textrm{g}(\Delta_\textrm{g},\Delta_\textrm{l}) $ into
the point  $\mu^{*}=0$, $\Delta_\textrm{g}=-d/2$, and the
product $12G\Delta_\textrm{g}\Delta_\textrm{l}^2$ nullifies the liquid branch in the
point $\mu^{*}=0, \,\Delta=d/2$, and we get the curve presented in figure~\ref{fig:8}.

It remains to clarify the problem of the horizontal branch in the
equation of state, the fact of originating of the liquid phase in the
gaseous one at isothermal compression of the gas, as well as the form
of the left-hand side in equation~(\ref{eq3.17}) for the equation of state.

This way, the existence of mixed products $\Delta_\textrm{g}^m\Delta_\textrm{l}^n$
in equations~(\ref{eq4.7}), (\ref{eq4.8}) leads to the appearance of a shelf
on the curve $ \mu^{*}(\Delta) = \mu^{*}_\textrm{g}(\Delta_\textrm{g},\Delta_\textrm{l})+\mu^{*}_\textrm{g}(\Delta_\textrm{g},\Delta_\textrm{l}) $
in the equation of state. Conditions~(\ref{eq4.6}) have a fundamental significance because
condensation into the liquid starts at density of gas $\eta_\textrm{g}=\eta_\textrm{c}(1-d/2)$, and terminates at
density of liquid $\eta_\textrm{l}=\eta_\textrm{c}(1+d/2)$. In other words, the gas isotherm ends at the
point $ \rho_{02}=d,\,\Delta_\textrm{g}=-d/2 $ at density $\eta_\textrm{g}=\eta_\textrm{c}(1-d/2)$, and the
liquid-phase isotherm ends at the point $\rho_{01}=d, \, \Delta_\textrm{l} = d/2$ at
density $\eta_\textrm{l}=\eta_\textrm{c}(1+d/2)$. On the line $ \eta_\textrm{c}(1-d/2) \div \eta_\textrm{c}(1+d/2)$,
the phase transition occurs from the gas into liquid. This phase transition from
gas into liquid along $ \mu_\textrm{g}^*(-d/2;d/2) \div \mu_\textrm{l}^*(-d/2;d/2) $ occurs due to
the external latent work of the pressure.

Therefore, according to~(\ref{eq4.7}), there are contributions into the energy
density of the surface layer due to $ \mu_\textrm{g}^*(\Delta_\textrm{g},\Delta_\textrm{l}) $ and $ \mu_\textrm{l}^*(\Delta_\textrm{g},\Delta_\textrm{l}) $
\bea
\label{eq4.11}
&& \left[\mu^{*}_\textrm{g}(\Delta_\textrm{g},\Delta_\textrm{l})\right]_{\rm surf} = 12G\Delta_\textrm{g}\Delta_\textrm{l}^2, \nonumber \\
&& \left[\mu^{*}_\textrm{l}(\Delta_\textrm{g},\Delta_\textrm{l})\right]_{\rm surf} = 12G\Delta_\textrm{g}\Delta_\textrm{l}^2,
\\
&& \mu^{*}_\textrm{l}(-d/2,d/2)-\mu^{*}_\textrm{g}(-d/2,d/2) = -\frac{3}{2}Dd.
\eea
Since $\Delta_\textrm{g}=w_\textrm{g}\rho_{02}$, $\Delta_\textrm{l}=w_\textrm{l}\rho_{01}$, and the
probabilities $ w_\textrm{g} $ and $ w_\textrm{l} $ depend on the number of particles $N$
and the angle $\varphi$, as is seen in figure~\ref{fig:8}. At $N$ increasing,
the products of $ w_\textrm{g} $ and $ w_\textrm{l} $ degenerate into $\delta-$functions
placed along $\varphi = \pi/2$. This means that in the limit of large $N$, the
two-phase liquid-gas system degenerates into the system gas~--- a droplet of
liquid divided by the surface without particles.

The energy of the phase layer transforms into the energy associated with the creation
of a geometric surface of division. The two-phase system emerges: gas and a
droplet of liquid of radius $R_\textrm{dr}$. If for the gaseous phase $P_\textrm{g}$ is the
pressure, $V_\textrm{g}$ is the volume, $P_\textrm{g}V_\textrm{g}$ is the energy, then for the droplet of
liquid the volume is $\frac{4\pi}{3}R_\textrm{dr}^3$, the pressure inside it
$P_\textrm{l} = \left(P_\textrm{g}+{2\alpha}/{R_\textrm{dr}}\right)$, where $\alpha$ is
the surface tension coefficient, the energy
$P_\textrm{l} V_\textrm{l} = \left(P_\textrm{g}+{2\alpha}/{R_\textrm{dr}}\right)\left({4\pi R_\textrm{dr}^3}/{3}\right)$ and plus
the surface tension energy $-\alpha S_\textrm{dr} = -\alpha 4\pi R_\textrm{dr}^2$.

Hence, the formation of the droplet with a geometric interface is concerned with an additional energy of the surface
\be
\label{eq4.12}
\frac{2\alpha}{R_\textrm{dr}} \frac{4\pi R_\textrm{dr}^3}{3} - \alpha4\pi R^2 = -\frac{\alpha}{R_\textrm{dr}}\frac{4\pi R_\textrm{dr}^3}{3}\,,
\ee
where the surface-energy density is $-{\alpha}/{R_\textrm{dr}}$. The magnitude of the
surface energy is connected with quantity~(\ref{eq4.11}). Also, the change of the volume
for the gaseous phase takes place, from $V_\textrm{l}$ to $V_\textrm{g}=V-({4\pi R^3}/{3})({n_\textrm{l}}/{n_\textrm{g}})$.

\section{The equality of the chemical potentials of gas and liquid on the \\ condensation-boiling line $\mu^{*}(\Delta_\textrm{g},\Delta_\textrm{l})$}
\label{sec:chem_equiv}
For $\mu^*$, by definition, we have equation~(\ref{eq9}):
\be
\label{eq5.1}
\mu^* = \beta (\mu - \mu_0) + \xi - |\alpha(0)|(1 - \Delta) .
\ee
In the approximation accepted in this research, the chemical potential of the reference system is
\be
\label{eq5.2}
\beta \mu_0 = \xi .
\ee
Then, for the chemical potential $\mu$ we get the following expression:
\begin{align}
\label{eq5.3}
 \mu &= - \Theta \, |\alpha(0)|(1 - \Delta), \nonumber\\
 \Delta &= w_{\textrm{g}} \rho_{02} + w_{\textrm{l}} \rho_{01}\, .
\end{align}
On the line of phase transition the conditions~(\ref{eq4.5}),~(\ref{eq4.6}), $\mu^{*}(\Delta_\textrm{g},\Delta_\textrm{l})=0$, hold together with
\be
\label{eq5.5}
N_{\textrm{g}} + N_{\textrm{l}} = N, \qquad \rd N_{\textrm{g}} = - \rd N_{\textrm{l}}\,,
\ee
where $N_{\textrm{g}}$ and $N_{\textrm{l}}$ are the number of particles in the gas phase
and in the liquid phase, respectively, $\mu_{\textrm{g}}$, $\mu_{\textrm{l}}$ are the chemical
potentials of the liquid and of the gas, respectively. According to~(\ref{eq4.6})
\begin{align}
\label{eq5.6}
\beta\mu_{\textrm{g}} &= |\alpha(0)|_{\textrm{g}}  (1 - \Delta_{\textrm{g}}), \nonumber\\
\beta\mu_{\textrm{l}} &= |\alpha(0)|_{\textrm{l}}  (1 - \Delta_{\textrm{l}}) .
\end{align}

According to definition~(\ref{eq12})
\begin{align}
\label{eq5.7}
 |\alpha(0)|\; &= \frac{N}{V} \frac{1}{\Theta} |\tilde \Phi (0)|, & &\text{then} \nonumber\\
 |\alpha(0)|_{\textrm{g}} &= \eta_{\textrm{g}} \frac{6}{\pi\sigma^3} \frac{1}{\Theta} |\tilde \Phi (0)|, & &
 \eta_{\textrm{g}}  \text{ --- gas density}, \nonumber\\
 |\alpha(0)|_{\textrm{l}} \,&= \eta_{\textrm{l}} \frac{6}{\pi\sigma^3} \frac{1}{\Theta} |\tilde \Phi (0)|, & & \eta_{\textrm{l}} \text{ --- liquid density in a droplet}.
\end{align}
We consider the beginning of  condensation (boiling)
\begin{align}
\label{eq5.8}
\Delta \ &= \Delta_{\textrm{g}} + \Delta_{\textrm{l}}\,, \nonumber\\
\Delta_{\textrm{g}} &= \frac12 \rho_{02}\,, \qquad  \rho_{02} = -d, \qquad \Delta_{\textrm{g}} = -\frac12 d,   \nonumber\\
\Delta_{\textrm{l}}\, &= \frac12 \rho_{01}\,,\qquad \rho_{01} = d, \qquad \phantom{-}\Delta_{\textrm{l}} = \frac12 d.
\end{align}
The densities $\eta_{\textrm{g}}$ and $\eta_{\textrm{l}}$ are expressed via the value
$\eta_{\textrm{c}}$ in the critical point and via $\Delta$ based on the relationship~(\ref{eq1.15}):
\be
\label{eq5.9}
\eta = \eta_\textrm{c} \Delta + \eta_\textrm{c}\,, \qquad \eta_{\textrm{g}} = \eta_\textrm{c} \Bigl(1 - \frac12 d \Bigr)\,, \qquad
\eta_{\textrm{l}} = \eta_\textrm{c} \Bigl(1 + \frac12 d \Bigr) \,.
\ee
Then, for~(\ref{eq4.4}) we obtain
\bea
\label{eq5.10}
&& \mu_{\textrm{g}} =  |\tilde \Phi(0)| \frac{6}{\pi\sigma^3} \eta_\textrm{c} \Bigl(1 - \frac{d^2}{4}\Bigr), \nonumber\\
&& \mu_{\textrm{l}} \,=  |\tilde \Phi(0)| \frac{6}{\pi\sigma^3} \eta_\textrm{c} \Bigl(1 - \frac{d^2}{4}\Bigr).
\eea
Thus,
\be
\label{eq5.11}
\mu_{\textrm{g}} = \mu_{\textrm{l}}\,.
\ee
It was proved that on the line $\mu^* = 0$ there exists a two-phase system of a
gas phase and of a liquid phase: the maternal sphere of the gas and the droplet
of liquid inside it  and the chemical potentials of both phases are the same. Let us
imagine that conditions~(\ref{eq5.8}) and~(\ref{eq5.9}) necessary for equalities~(\ref{eq5.10})
and~(\ref{eq5.11}), may be also valid  for other quantities $\Delta_{\textrm{g}}$ and $\Delta_{\textrm{l}}$,
not only for that of~(\ref{eq5.8}), which are located symmetrically to the point of the intersection
of the rectlinear diameter with the line $\mu^* =0$ that is $\Delta_{\textrm{g}} = - \Delta_{\textrm{l}}$.
Such points are located at the ends of the intervals which pass through the origin $\mu^{*}=0$ and touch the
lines $\rho_{02}$ and $\rho_{01}$ that are the curves of the maxima for gaseous and liquid phases,
respectively, see figure~\ref{fig:mu_w}. For each pair of points, denoted by $\tilde{\rho}_{02}$
and $\tilde{\rho}_{01}$, the equalities $E(\tilde{\rho}_{02}) = E(\tilde{\rho}_{01})$
take place, since $\tilde{\rho}_{02}=-\tilde{\rho}_{01}$, and $E(\rho_0)$ is an even function.
Thus, for each pair of end-point values there is $w(\tilde{\rho}_{02})=w(\tilde{\rho}_{01})$.
Hence, for each pair, the following is true:
$ w(\tilde{\rho}_{02})\tilde{\rho}_{02} =- w(\tilde{\rho}_{01})\tilde{\rho}_{01} $ and
$\tilde{\Delta} = w(\tilde{\rho}_{02})\tilde{\rho}_{02} + w(\tilde{\rho}_{01})\tilde{\rho}_{01}=0$, $\mu^{*}=0$, ${\cal{E}}(0)=0$,
and also $\mu_\textrm{g}(\tilde{\Delta}_\textrm{g})=\mu_\textrm{l}(\tilde{\Delta}_\textrm{l})$. However, conditions~(\ref{eq4.6}) are not obeyed if $\Delta_\textrm{g}=-d/2$,
$\Delta_\textrm{l}=d/2$ are substituted by any of $\tilde{\Delta}_\textrm{g}=-\tilde{\Delta}_\textrm{l}$.

Therefore, neither approaching nor moving away from the densities of gas and liquid
compared to~(\ref{eq5.8}) occur. The conclusion is that densities for gaseous
and liquid phases do not change at the first-order phase transtition. The density of gas remains fixed
\be
\label{eq5.6a}
\eta_{\textrm{g}} = \eta_\textrm{c} \Delta_{\textrm{g}} + \eta_\textrm{c} = \eta_\textrm{c} \Bigl( 1 - d/2 \Bigr).
\ee
The density of liquid remains fixed
\be
\label{eq5.7a}
\eta_{\textrm{l}} = \eta_\textrm{c} \Delta_{\textrm{l}} + \eta_\textrm{c} = \eta_\textrm{c} \Bigl( 1 + d/2 \Bigr).
\ee
The chemical potentials of both phases are the same
\[
\mu_{\textrm{g}} = \mu_{\textrm{l}} .
\]
This remains true as long as the phase transition takes place. As a result,
under the action of latent work of the pressure, all the gas condenses into
liquid. A jump-like change of the gas density takes place
\be
\label{eq5.8a}
\eta_{\textrm{l}} - \eta_{\textrm{g}} = \eta_\textrm{c} d .
\ee

\section{The process of condensation of gas into liquid (the boiling of liquid)}
\label{sec:condens_proc}

This section begins with~(\ref{eq3.17})
\be
\label{eq6.1}
\frac{PV}{\Theta} = N{\cal{E}}(\Delta), \qquad P=\tilde{\Theta}\eta{{\cal{E}}}(\Delta), \qquad \tilde{\Theta}=\Theta\frac{6}{\pi\sigma^3}\,.
\ee
Let us follow the process of isothermal quasistatic compression in the
gaseous phase (see section~\ref{sec:intro}). The process takes place along the line
$\Delta = w_{\textrm{g}}\rho_{02} + w_{\textrm{l}}\rho_{01}$. Starting
at $\Delta = \Delta_\textrm{g}=w_\textrm{g}\tilde{\rho_{02}}$:
\be
\label{eq6.2}
\rho_{02} = -b, \qquad w_{\textrm{g}} = 1, \qquad \Delta = -b, \qquad w_{\textrm{l}} = 0,
\ee
where (see figures~\ref{fig:mu_w}--\ref{fig:Delta}) there is a homogeneous
gaseous system (see figures~\ref{fig:mu_w} and~\ref{fig:three_quan}).

Next, after the quantities $\rho_{02}$ and $\Delta$ increase, we get
$\rho_{02} > -b$, $\Delta > -b$, and the probabilities are as follows:
\[
w_{\textrm{g}} < 1, \qquad w_{\textrm{l}} > 0, \qquad w_{\textrm{g}} + w_{\textrm{l}} = 1.
\]
In a gaseous system, there emerge density fluctuations of a liquid type.
During the quasistatic process for $-b \leqslant \rho_{02} \leqslant ff_{\textrm{l}}$
fluctuations are unstable, because, according to figure~\ref{fig:three_quan}, under
the line $ff_{\textrm{l}} - f_{\textrm{l}}$ $E(\Delta) < 0$ and $\exp [NE(\Delta)] \to 0$.

Being further compressed, while crossing the line $ff_{\textrm{l}} - f_{\textrm{l}}$
(see figure~\ref{fig:three_quan}) in a gaseous system, there emerge stable density
fluctuations of a liquid type because  $E(\Delta) > 0$ therein. The
argument $\Delta$ is determined by equation~(\ref{eq3.10}). Here, both terms
in $\Delta$ become of non-zero value. However, the term $w_{\textrm{g}}\rho_{02}$
in $\Delta$ is a principal one. As earlier, we have only gas, and there is a single-phase system (see figure~\ref{fig:Delta}).

When the external pressure increases, the pressure in the gaseous system
also increases. The density of a gaseous system grows, in accordance
with~(\ref{eq1.15}). The gaseous system evolves along the same isotherm $\mu^{*}_\textrm{g}(\rho_{02})$.
Now we get to the limiting densities $\eta_\textrm{g}=\eta_\textrm{c}(1+w_\textrm{g}\rho_{02})$, $\eta_l=\eta_\textrm{c}(1+w_l\rho_{01})$, where
\be
\label{eq6.3}
\rho_{02} = - d,\qquad \rho_{01} =  d, \qquad E(d) = E(-d), \qquad w_{\textrm{g}} = w_{\textrm{l}} = 1/2.
\ee
Based on (\ref{eq3.10}), (\ref{eq4.6}), (\ref{eq4.8})--(\ref{eq4.10}), it follows
\be
\label{eq6.4}
\Delta = \Delta_\textrm{g}+ \Delta_\textrm{l} = 0, \qquad \mu^{*} = \mu^{*}_\textrm{g} + \mu^{*}_\textrm{l} = 0, \qquad
{\cal{E}}={\cal{E}}_\textrm{g}+{\cal{E}}_\textrm{l}=0.
\ee
This state describes the end of the curve for the principal maxima for a
gaseous phase and the beginning of the principal maxima for a liquid phase.
Here, we arrive at~(\ref{eq4.10}) and~(\ref{eq6.1}).

When work is done on the gaseous system, within the gas system
there  arise processes that evolve along the curve $\Delta_{\textrm{l}} = w_{\textrm{l}}\rho_{01}$,
although under the conditions of a secondary maximum, as is seen in figures~\ref{fig:mu_w} and~\ref{fig:three_quan}.
The system does an internal work to create stable density fluctuations of liquid type,
which in the case of $ \rho_{02}=-d$, $\Delta_\textrm{g} = -d/2$, $\rho_{01} = d$, $\Delta_\textrm{l}=d/2 $
causes the creation of a droplet.

In order to determine the properties of a droplet, it should have exact boundaries.
Equations~(\ref{eq4.7}) and~(\ref{eq4.8}) for $ \mu^*$, $\mu^*_\textrm{g}$, $\mu^*_\textrm{l}$, ${\cal{E}}$,
${\cal{E}}_\textrm{g}$, ${\cal{E}}_\textrm{l}$
are ``bulk'' formulas. In these formulas, the interface is a phase layer with a number of particles.
Our further calculations  are performed for the case of very large $N$. The model of a
two-phase region is considered as gas and a droplet of liquid arising in gas, the
interface is a geometric surface without particles. The mixed products $ \Delta_\textrm{g}^m \Delta_\textrm{l}^n$,
entering $ \mu^{*}(\Delta_\textrm{g},\Delta_\textrm{l})$ and $ {\cal{E}}(\Delta_\textrm{g},\Delta_\textrm{l})$,
presented in~(\ref{eq4.7}) and~(\ref{eq4.8}) are replaced with the energy of additional
pressure in the droplet of liquid $({4\pi R_\textrm{dr}^3}/{3})({2\alpha}/{R_\textrm{dr}})$
and the surface energy $-\alpha4\pi R_\textrm{dr}^2$.

The two-phase system exists only along the line $\mu^* =0$. On this line, the chemical
potentials of gaseous and liquid phases are the same. Equality~(\ref{eq4.11}) is invariant
of the process of condensation (i.e., boiling). This is true for the case when the thermodynamic
potential $\Phi_{\rho_0}$ from the beginning to the end of the process of condensation
(i.e., boiling) remains constant
\be
\label{eq6.5}
\Phi_{\rho_0} = N_{\textrm{g}} \mu_{\textrm{g}} + N_{\textrm{l}} \mu_{\textrm{l}} = N \mu_{\textrm{g}} =
N \mu_{\textrm{l}}\,,
\ee
since $N_{\textrm{g}} + N_{\textrm{l}} = N$ remains constant and is equal to the initial
number of particles, and $\rd N_{\textrm{g}} = -\rd N_{\textrm{l}}$, which, in fact, holds.
The amount of gas decreases, while the amount of liquid increases till all the gas transforms into liquid.

Liquid phase originates in the form of a droplet confined by the interface $S$ of some surface energy.

The sharp bend of the isotherm in figure~\ref{fig:8} is related to the instant of the
creation of a droplet, and the sum of the moduli of the mixed products~(\ref{eq4.11}) is equal to the energy of the droplet.

\subsection{The energy of the surface tension}
 According to~(\ref{eq3.10}),~(\ref{eq5.6})--(\ref{eq5.8}), along the line $\mu^* =0$ we have,
$
\Delta = \Delta_{\textrm{g}} + \Delta_{\textrm{l}}$,
$\Delta_{\textrm{g}} = w_{\textrm{g}} (-d) = - d/2$,
$\Delta_{\textrm{l}} = w_{\textrm{l}} d =  d/2$, $w_{\textrm{g}} = w_{\textrm{l}} = 1/2
$.

For the basic relationship~(\ref{eq6.1}), the left-hand side takes the form
\be
\label{eq6.5a}
P_{\textrm{g}} V_{\textrm{g}} + P_{\textrm{l}} V_{\textrm{l}} - \alpha S_{\textrm{l}}\,,
\ee
where $P_{\textrm{g}}$, $P_{\textrm{l}}$ are the pressures in the gas and
in the liquid phases, $V_{\textrm{g}}$ and $V_{\textrm{l}}$ are the volumes
of the gas and of the liquid phases, $S_\textrm{l}$ is the surface of the droplet of
liquid, $\alpha$ is the surface-tension coefficient. In the right-hand side of~(\ref{eq6.1}) we obtain:
\be
\label{eq6.6}
N_{\textrm{g}}{\cal{E}}_{\textrm{g}}(\Delta_{\textrm{g}}) + N_{\textrm{l}}{\cal{E}}_{\textrm{l}}(\Delta_{\textrm{l}}) \simeq
N_{\textrm{g}}\mu^{*}_{\textrm{g}}(\Delta_{\textrm{g}}) + N_{\textrm{l}}\mu^{*}_{\textrm{l}}(\Delta_{\textrm{l}}).
\ee
According to the conclusions drawn in~(\ref{eq5.6}) and~(\ref{eq5.7}),
the quantities $\mu^{*}_\textrm{g}, \mu^{*}_\textrm{l}$ take on fixed values.
In accordance with~(\ref{eq5.6})--(\ref{eq5.8}):
\bea
\label{eq6.7}
&& {\cal E}_{\textrm{g}}(\Delta_{\textrm{g}}) = \mu^*(\Delta_{\textrm{g}}) = -2D\Delta_{\textrm{g}} + 4G\Delta_{\textrm{g}}^3 = 3Dd/4,\nonumber\\
&& {\cal E}_{\textrm{l}}(\Delta_{\textrm{l}}) = \mu^*(\Delta_{\textrm{l}}) = -2D\Delta_{\textrm{l}} + 4G\Delta_{\textrm{l}}^3 =-3Dd/4 ,
\eea
and the process of the phase transition is reduced to redistribution of $N_\textrm{g}$
and $N_\textrm{l}$. The width of the interface between gas and liqud droplet is considered to be zero.

Liquid phase is a droplet of radius $R$ that originates in the gaseous phase.
The droplet has strict boundaries. Combining~(\ref{eq6.5})--(\ref{eq6.7}), for~(\ref{eq6.1}) on the line $\mu^* = 0$ we obtain:
\be
\label{eq6.8}
P_{\textrm{g}} V_{\textrm{g}} + P_{\textrm{l}} V_{\textrm{l}} - \alpha S_{\textrm{l}} = \Theta\{ N_{\textrm{g}} {\cal E}_{\textrm{g}} + N_{\textrm{l}} {\cal E}_{\textrm{l}} \}.
\ee
The emergence of a liquid droplet, restricted by a shell, in a gaseous phase
is the essence of the first order phase transition. The droplet is of a spherical
form. The surface tension energy is minimal. For~(\ref{eq6.8}), we get:
\bea
\label{eq6.9}
&& P_{\textrm{g}} V_{\textrm{g}} = P_{\textrm{g}} V - P_{\textrm{g}} \frac{4\pi}{3} R_{\textrm{dr}}^3 \frac{\eta_{\textrm{l}}}{\eta_{\textrm{g}}}\,,   \nonumber\\
&& P_{\textrm{l}} V_{\textrm{l}} = \left( P_{\textrm{g}} + \frac{2\alpha}{R_{\textrm{dr}}} \right) \frac{4\pi}{3} R_{\textrm{dr}}^3 \,,
\eea
where $V$ is the initial volume, $R_{\textrm{dr}}$ is the radius of the
first droplet, $\eta_{\textrm{l}}$, $\eta_{\textrm{g}}$ are the densities of gas and liquid, respectively.
For the energy of the droplet surface, we have:
\be
\label{eq6.10}
- \alpha S = - \alpha 4\pi R_{\textrm{dr}}^2 = - \frac{3\alpha}{R_{\textrm{dr}}} \frac{4\pi R_{\textrm{dr}}^3}{3} \,.
\ee

In the right-hand side of~(\ref{eq6.8}), it is written
\be
\label{eq6.11}
\Theta [N_{\textrm{g}} {\cal E}_{\textrm{g}} + N_{\textrm{l}} {\cal E}_{\textrm{l}} ] =
\Theta [N {\cal E}_{\textrm{g}} + N_{\textrm{l}} ({\cal E}_{\textrm{l}} - {\cal E}_{\textrm{g}})] .
\ee
Collecting~(\ref{eq6.7})--(\ref{eq6.9}) in equation~(\ref{eq6.6}), we obtain:
\bea
\label{eq6.12}
&& P_{\textrm{g}} V_{\textrm{g}} + P_{\textrm{l}} V_{\textrm{l}} - \alpha S = P_{\textrm{g}} V -  \frac{\alpha}{R_{\textrm{dr}}} \frac{4\pi R_{\textrm{dr}}^3}{3} - P_{\textrm{g}}  \frac{4\pi R_{\textrm{dr}}^3}{3} \left(  \frac{\eta_{\textrm{l}}}{\eta_{\textrm{g}}} -1 \right),   \nonumber\\
&& P_{\textrm{g}} \left[ V - \frac{4\pi R_{\textrm{dr}}^3}{3} \left(  \frac{\eta_{\textrm{l}}}{\eta_{\textrm{g}}} - 1 \right) \right] - \frac{\alpha}{R_{\textrm{dr}}} \frac{4\pi R_{\textrm{dr}}^3}{3} = N\Theta {\cal E}_{\textrm{g}} + N_{\textrm{l}} ({\cal E}_{\textrm{l}} - {\cal E}_{\textrm{g}}) .
\eea
The change of gas volume by
$\frac{4\pi}{3}R_{\textrm{dr}}^3 \left(  {\eta_{\textrm{l}}}/{\eta_{\textrm{g}}} - 1 \right) $,
that takes place in the product $P_{\textrm{g}} \left[ V - \frac{4\pi }{3}R_0^3
\left({\eta_{\textrm{l}}}/{\eta_{\textrm{g}}} - 1 \right) \right]$ in (\ref{eq6.12})
corresponds to those changes in the quantity ${\cal E}_{\textrm{g}}(\Delta_{\textrm{g}})$
which were observed at the transition from ${\cal E}_{\textrm{g}}(\Delta_{\textrm{g}})$
at $\Delta_{\textrm{g}} = -b$ to ${\cal E}_{\textrm{g}}(\Delta_{\textrm{g}})$
at $\Delta_{\textrm{g}} = w_{\textrm{g}}(-d) = -d/2$. We compare
\be
\label{eq6.13}
P_{\textrm{g}} \left[ V - \frac{4\pi R^3}{3} \left(  \frac{\eta_{\textrm{l}}}{\eta_{\textrm{g}}} - 1 \right) \right] = N\Theta   {\cal E}_{\textrm{g}} \,.
\ee
Hence, for the specific\footnote{By a volume unity.} surface tension energy of the droplet of radius $R$, we have
\be
\label{eq6.12a}
- \frac{\alpha}{R} = n_{\textrm{l}} ( {\cal E}_{\textrm{l}} - {\cal E}_{\textrm{g}})\Theta .
\ee
Let us replace ${\cal E}_{\textrm{l}}$ and ${\cal E}_{\textrm{g}}$ with~(\ref{eq6.7}),
and  $n_{\textrm{l}}$ with $n_{\textrm{l}} = N_{\textrm{l}}/ ({4\pi R_{\textrm{dr}}^3}/{3})
= \eta_{\textrm{l}} {6}/{\pi \sigma^3}$. In the right-hand side of (\ref{eq6.12a}), we obtain:
\be
\label{eq6.13a}
\Theta n_{\textrm{l}} ( {\cal E}_{\textrm{l}} - {\cal E}_{\textrm{g}}) = - \eta_{\textrm{l}} \frac{9\Theta}{\pi \sigma^3}Dd ,
\ee
which conforms to~(\ref{eq4.10}). Therefore, based on~(\ref{eq6.12a}),
there is a contribution to specific pressure $P_{\textrm{total}}$ from the surface tension energy of the liquid droplet of radius $R$:
\be
\label{eq6.13b}
P_{\textrm{total}} = -\frac{\alpha}{R} = -\frac{3}{2}\Theta\frac{6}{\pi \sigma^3}\eta_{\textrm{l}} D d = -\frac{3}{2}{\tilde \Theta}
\eta_{\textrm{l}} D d ,
\ee
where ${\tilde \Theta} = \Theta~6/\pi \sigma^3$.

The quantity $P_{\textrm{total}}$ is proportional to $\tau^{5/2\nu}$.
It governs the rate of decreasing the surface tension energy at approximatiion
to the critical point $\tau = 0$.

\subsection{Latent work of condensation (latent heat of boiling)}

Condensation begins at the moment of the creation of the droplet with the
surface energy~(\ref{eq6.13b}). As was mentioned above, condensation of a gas starts from the state
\[
N{\cal{E}}(\Delta)=N\mu^{*}_\textrm{g}(\Delta_\textrm{g}), \qquad \rho_{02} = -d, \qquad \Delta_{\textrm{g}} = -\frac12 d, \qquad \eta_{\textrm{g}} = \eta_\textrm{c} - \frac12 d \eta_\textrm{c}
\]
and terminates in a liquid system in the state
$ N{\cal{E}}(\Delta) = N\mu^{*}_\textrm{l}(\Delta_\textrm{l})$, $\rho_{01} = d$, $\Delta_{\textrm{l}} = \frac12 d $
and $\eta_{\textrm{l}} = \eta_\textrm{c} + \frac12 d \eta_\textrm{c}$. At the moment when condensation starts,
the pressure is equal to
\be
\label{eq6.14}
P_{\textrm{g}} = n_{\textrm{g}} \Theta {\cal E}_{\textrm{g}} (\Delta_{\textrm{g}}) = n_{\textrm{g}} \Theta \mu^* (\Delta_{\textrm{g}}).
\ee
Replacing $\Delta_{\textrm{g}}$ with $\Delta_{\textrm{g}} = w_{\textrm{g}} (-d) = - \frac12 d$ and
\be
\label{eq6.15}
\mu^* (\Delta_{\textrm{g}}) = dD - \frac12 d^3 G = \frac34 Dd .
\ee
Then,
\be
\label{eq6.16}
P_{\textrm{g}} = n_{\textrm{g}} \Theta \frac34 Dd .
\ee

The initial volume of gas is a sphere of radius $R_0$, of volume $V_0 = {4\pi R_0^3}/{3}$.
The final volume of gas after transferring into a liquid during condensation at a fixed pressure
is the volume of liquid $V_{\textrm{l}}$ in the form of a sphere of radius $R_{\textrm{l}}$:
\[
V_{\textrm{l}} = \frac{4\pi}{3} R_{\textrm{l}}^3 .
\]
All $N$ particles of a gaseous sphere $V_0$ passed into a sphere of liquid $V_{\textrm{l}}$. Thus,
\[
\frac{1}{v_{\textrm{g}}} V_0 = \frac{1}{v_{\textrm{l}}}V_{\textrm{l}}\,, \qquad
V_{\textrm{l}} = \frac{v_{\textrm{l}}}{v_{\textrm{g}}}V_0 \,,
\]
the total change of the volume is
\be
\label{eq6.17}
(V_{\textrm{l}} - V_0) = \left( \frac{v_{\textrm{l}}}{v_{\textrm{g}}} - 1 \right) V_0 \,.
\ee
The work performed is the latent work (i.e., heat) of the process of condensation:
\[
- P_{\textrm{g}} (V_{\textrm{l}} - V_0) = - P_{\textrm{g}} \left(\frac{v_{\textrm{l}}}{v_{\textrm{g}}} - 1 \right) \frac{4\pi}{3} R_\textrm{l}^3 \,.
\]

Having substituted~(\ref{eq6.16}), the specific latent work of
condensation is\footnote{Dimension of $L$ coincides with the dimension of pressure.}
\be
\label{eq6.18}
L = - \left( 1 - \frac{v_{\textrm{l}}}{v_{\textrm{g}}} \right) n_{\textrm{g}} \Theta \frac34 D d = - (v_{\textrm{g}} - v_{\textrm{l}}) \frac34 \Theta D d n_{\textrm{g}}^2 \,.
\ee

Condensation at isothermal compressibility is a dynamical process.
Equation~(\ref{eq6.8}) concerns the states of a system in which
two phases have already been created: the initial gaseous phase and a new liquid phase in the form of a droplet. The conditions $\Delta = 0$, $\mu^* = 0$, ${\cal E} (\Delta) =0$
require an external pressure to be applied in order to do the latent work of condensation.
As a result, the gaseous phase disappears (or remains as a nascent bubble of gas), and we write:
\be
\label{eq6.19}
P_{\textrm{l}} V_{\textrm{l}} = L (V_{\textrm{g}} - V_{\textrm{l}}) .
\ee
[For a nascent bubble of gas, if any, $P_{\textrm{g}} V_{\textrm{g}}^{~\textrm{nascent}}
- \alpha S_{\textrm{g}}^{~\textrm{nascent}}$ is energy, and we can write: $P_{\textrm{l}} V_{\textrm{l}}
+ P_{\textrm{g}} V_{\textrm{g}}^{~\textrm{nascent}} - \alpha S_{\textrm{g}}^{~\textrm{nascent}}
= L (V_{\textrm{g}} - V_{\textrm{g}}^{~\textrm{nascent}} - V_{\textrm{l}})$].

For the two-phase system along the line $\mu^* = 0$, ${\cal E}(\Delta) = 0$ for external pressure we write down:
\be
\label{eq6.20}
P_{\mu^* = 0} =  L \Bigl( \frac{v_{\textrm{g}}}{v_{\textrm{l}}} - 1 \Bigr) .
\ee
Using $L$, we get
\be
\label{eq6.21}
P_{\mu^* = 0} = \left( 1 - \frac{v_{\textrm{l}}}{v_{\textrm{g}}} \right) n_{\textrm{g}} \frac34 \Theta       D d .
\ee
On the curve of the equation of state, this expression represents a horizontal fragment
of the transition from a gaseous phase into a liquid phase under the isothermal
compression of a gaseous phase.

\section{The equation of state}
\label{sec:state_equation}
Now we shall summarize the results of our calculations in the critical region of
$\tau < 0$, $|\tau| < \tau^*$ and $|\eta - \eta_\textrm{c}| < |\eta^* - \eta_\textrm{c}|$, where
$\tau^*$ and $|\eta^* - \eta_\textrm{c}|$ are the boundaries of the critical region presented
in expressions~(\ref{eq15}),~(\ref{eq16}) and~(\ref{eq1.13}).

The condensation of gas at isothermal compression is a process analogous to the
boiling at an isothermal decrease of pressure of a liquid. This is a quasiequilibrial
directed process. Equation~(\ref{eq6.8}) concerns the states of a system in which two phases
have already been formed: the initial gaseous phase and a new liquid one, in the form of a droplet.
(The droplet has got a shell that separates liquid from gas, and its creation is the essence of
the first-order phase transition.) Conditions $\Delta = 0$, $\mu^* = 0$, $ {\cal{E}}(\Delta)=0$
require an external pressure to be applied in order to do the latent work of condensation (\ref{eq6.18}).
As a result, in (\ref{eq6.8}) the gaseous phase disappears (or remains as a nascent bubble of gas).

The conditions existing on the line of condensation $\mu^{*}=0$, ${\cal{E}}(\Delta)=0$,
$\Delta_\textrm{g}=-d/2 \div \Delta_\textrm{l}=d/2$, $\eta_\textrm{g} = \eta_\textrm{c}(1-d/2)\div \eta_\textrm{c}(1+d/2)=\eta_l$ are
determined by~(\ref{eq5.10}), i.e., by the equality of the chemical potentials of different phases.
The thermodynamic potential $\Phi_{\rho_0}$ of a system on the line of
condensation remains constant
\be
\Phi_{\rho_0} = N_\textrm{g}\mu_\textrm{g} +N_\textrm{l}\mu_\textrm{l} = N\mu_\textrm{g} = N\mu_\textrm{l}\,,
\ee
according to~(\ref{eq5.10})
\be
\label{eq7.1}
\Phi_{\rho_0} = N|\tilde{\Phi}(0)|\frac{6}{\pi\sigma^3}\eta_\textrm{c}(1-d^2/4).
\ee
On the line of the first-order phase transition, equations~(\ref{eq6.8}) hold.
To obtain these equations, the mixed products $\Delta_\textrm{g}^m\Delta_\textrm{l}^n$ were
nullified in the exact relationships~(\ref{eq4.7})--(\ref{eq4.8}), in particular
those in~(\ref{eq4.11}). Instead of them, the surface tension energy of a droplet
was introduced as well as an additional internal pressure of a droplet. This
allowed us to describe the magnitude of energy associated with the creation of the
liquid droplet. The droplet has got a strictly defined interface. We have described the quantity
of the latent work of the pressure.

\begin{figure}[!t]
	\centerline{
\includegraphics[angle=0,width=0.6\textwidth]{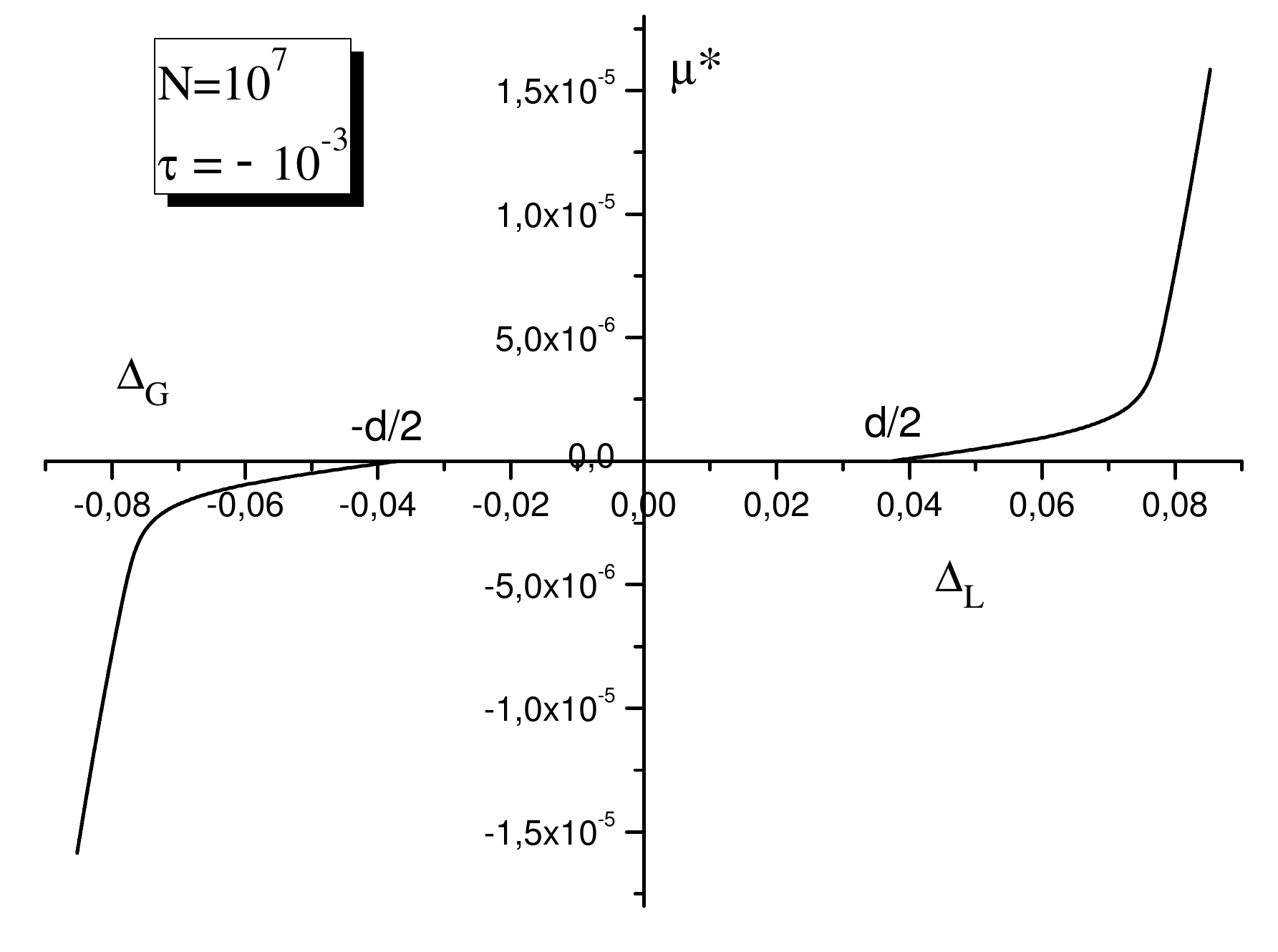}
}
	\caption{The graph of ${\cal E}(\Delta_{\textrm{g}}) \simeq \mu_{\textrm{g}}^*(\Delta_{\textrm{g}})$ and
		${\cal E}(\Delta_{\textrm{l}}) \simeq \mu_{\textrm{l}}^*(\Delta_{\textrm{l}})$;
circles denote densities of the gaseous phase $\eta_{\textrm{g}}^* = \eta_\textrm{c} (1 - d/2)$
and of the liquid phase $\eta_{\textrm{l}}^* = \eta_\textrm{c} (1 + d/2)$, which remains constant during
the liquid-gas phase transition. Both solutions are in the region of instability $\rho_{03} = b\cos
\frac{\varphi + 4\pi}{3}$ (see figures~\ref{fig:mu_w} and~\ref{fig:three_quan}). The jump of
the density at phase transition is $\eta_{\textrm{l}} - \eta_{\textrm{g}} = \eta_\textrm{c} d$. The
slope of the gaseous branch $\Delta_\textrm{g}$ while approaching the point $\Delta_\textrm{g}=-d/2$ is
equal to $0$. The slope of the liquid branch while turning away from the point $\Delta_\textrm{l}=d/2$
is equal to $0$. } \label{fig:mu_delta}
\end{figure}

The task of our further research will be to clarify the problem of the thermodynamic limit and to compare the
theory with experimental results. We will need
a total expression for $\ln{\Xi}$, not only $\ln{\Xi_{\rho_0}}$. These urgent problems
are easier to consider farther from the critical point, since there are no requirements
following from the renormalization-group symmetry near $T=T_\textrm{c}$.

Thus, the general expression for the equation of state takes the form
\bea
\label{eq7.2}
 P_{\rho_0} &=& \tilde \Theta \left\{ \eta_{\textrm{g}} \mu^*_\textrm{g}(\Delta_\textrm{g},\Delta_\textrm{l})_{\Delta_\textrm{g} \leqslant -d/2, \, \mu^* < -h/2} \right. \nonumber\\
&& + \; [\text{``shelf'',  the  region of constant pressure}]_{-h/2\leqslant \mu^{*}\leqslant h/2, \, -d/2\leqslant \Delta \leqslant d/2} \nonumber \\
&& + \left.  \eta_{\textrm{l}} \mu^*_\textrm{l}(\Delta_\textrm{g},\Delta_\textrm{l})_{\Delta \geqslant d, \, \mu^* > h/2} \right\},
\eea
where
\be
\mu^*_\textrm{l}(\Delta_\textrm{g},\Delta_\textrm{l}) = - 2 D\Delta_\textrm{l} + 4G \Delta_\textrm{l}^3 +12G\Delta_\textrm{l}\Delta_\textrm{g}^2 \,,\qquad h\to 0,\nonumber
\ee
\be
\mu^*_\textrm{g}(\Delta_\textrm{g},\Delta_\textrm{l}) = - 2 D\Delta_\textrm{g} + 4G \Delta_\textrm{g}^3 +12G\Delta_\textrm{g}\Delta_\textrm{l}^2\,, \qquad h\to 0. \nonumber
\ee
On the ``shelf'', the external pressure is equal to the specific work of condensation
according to~(\ref{eq6.8}), $P_{\mu^{*}}=\left(1-{v_\textrm{l}}/{v_\textrm{g}}\right)\eta_\textrm{g}\tilde{\Theta}\frac{3}{4}Dd$,
and the thermodynamic potential remains constant, $\Phi_{\rho_0} = N\mu_\textrm{g}(-d/2,d/2)=N\mu_\textrm{l}(-d/2,d/2)$.

The phase transition is a dynamical process. It takes place without any breaks as a
quasistatic isothermal compression of a gaseous phase. There are five stages of the process.
The first one is a gaseous phase with density $\eta=\eta_\textrm{c}(-b)+\eta_\textrm{c}$. Here,
$w_\textrm{g}(\varphi)=w_\textrm{g}(\pi)=1$, $w_\textrm{l}=0$, $\Delta = \Delta_\textrm{g}=-b$, $b=\sqrt{2D/(3G)}$, $\Delta_\textrm{l}=0$.
When moving along $\Delta=\Delta_\textrm{g}=w_\textrm{g}(\varphi)\rho_{02}(\varphi)$,
there are  density fluctuations of a liquid type in the gaseous state along the
secondary maximum $\Delta_\textrm{l}=w_\textrm{l}\rho_{01}$.
The region of a sharp bend of the isotherm $\mu^{*}_\textrm{g}(\Delta_\textrm{g},\Delta_\textrm{l})$
is reached due to the terms
in (\ref{eq4.11}) (see figure~\ref{fig:mu_delta}). In a gaseous phase, a liquid
droplet appears with an additional, in comparison with the gaseous system, energy
 density~(\ref{eq6.12a})--(\ref{eq6.14}). A liquid droplet that emerges means
 the end of the first stage and the beginning of the second stage, being the main
 stage of the first-order phase transition.

The thickness of interface, accordint to figure~\ref{fig:7a} is equal to half-width
of the maximum for the expressions $12\Delta_\textrm{g}\Delta_\textrm{l}^2$, $12\Delta_\textrm{l}\Delta_\textrm{g}^2$,
and $2\Delta_\textrm{g}\Delta_\textrm{l}$. In general, the interface includes all the mixed products
$2\Delta_\textrm{g}^m\Delta_\textrm{l}^n$, via $\mu^{*}(\Delta_\textrm{g},\Delta_\textrm{l})$ and ${\cal{E}}(\Delta_\textrm{g},\Delta_\textrm{l})$.
All terms of the mixed products distribute accordingly in equations~(\ref{eq4.7}) and~(\ref{eq4.8})
between ${\cal{E}}_\textrm{g}$ and ${\cal{E}}_\textrm{l}$ in such a way that in the fixed limiting points of a
phase transition $\Delta_\textrm{g}=-d/2$ and $\Delta_\textrm{l}=d/2$, the fundamental equalities~(\ref{eq4.6}) hold.

The third stage is considered to be the result of the action of the external latent
work of pressure on the two-phase system, presented in~(\ref{eq6.19})--(\ref{eq6.20}).
Pressure and temperature do not change, accoridng to~(\ref{eq5.6})--(\ref{eq5.7}).
Gas transfers into liquid according to~(\ref{eq6.5}). The external latent work of pressure
has been done. The density jump $\Delta\eta = \eta_\textrm{l}-\eta_\textrm{g} = \eta_\textrm{c}d$ takes place.
Everything ends with the fourth stage, which is a liquid with a nascent bubble of gas. Finally,
the fifth stage is a liquid of density growing from the value $\eta_\textrm{l}=\eta_\textrm{c}d/2
+\eta_\textrm{c}$ to $\eta_\textrm{l}=\eta_\textrm{c}b+\eta_\textrm{c}$. Thus, the events in the region $Q<0$
finish. The region of liquid densities begins, where the discriminant $Q>0$. This is a homogeneous liquid phase.

We started with an initial sphere of gas in a boundless system. By isothermal compression,
we reached a point of the first-order phase transition. Gas transferred into liquid, condensed into a droplet.
The quantities that describe the process were determined. We worked within the framework of
the grand canonical distribution for the Gibbs statistics. The successful solution of the problem
is ensured by the choice of a related phase space of collective variables, among which there is
a variable $\rho_0$. This variable is responsible for a description of the phase
transition and the density jump. Calculations are carried out using the quartic basic
measure density. The zero-approximation performed in this work provides good results
on both the critical indexes and the description of the mechanism of density jump at
the condensation of gas into liquid.

We have gone through all the events of the first-order phase transition below the critical point.
The curve for the equation of state has a form presented in figure~\ref{fig:mu_delta}.
A peculiar thing is the density jump between the points $\eta_{\textrm{g}} = \eta_\textrm{c} \left( 1 - d/2 \right)$
and $\eta_{\textrm{l}} = \eta_\textrm{c} \left( 1 + d/2 \right)$, where $d = \sqrt{ {D}/{2G}}$, where
$D$ and $G$ are the coefficients at the second and fourth power of $\rho_0$, respectively, in the initial
Hamiltonian $E(\rho_0)$.

Figure~\ref{fig:mu_w} resembles the Van der Waals isotherm. The theory put forward by a great
physicist~\cite{vdWaals_Nobel}, was the basis for numerous researches devoted to
the gas-liquid phase transition, including the present research.

At the same time, this work is not void  of numerous evident drawbacks. Among them we note:
\begin{itemize}
\item we worked in a very narrow regions of temperatures and densities, near the critical point;

\item when calculating the Jacobian of transition to the phase space of collective variables,
the statistical weight was introduced, the distribution over short-range interaction between
hard spheres. The cumulants of the Jacobian, $\mathfrak{M}_1$, $\mathfrak{M}_2$, $\mathfrak{M}_3$
and $\mathfrak{M}_4$, were taken at $k = 0$. This was connected with the fact that they have
wide ``shelves'' at $k = 0$. The cumulants were determined via compressibility of the
reference system. We neglected the terms proportional to $k^2$ , which were proportional
to the binary correlation function for the system of hard spheres;

\item we neglected the effect of integration over $\rho_{\bf k}$ for $k \neq 0$ on the
form of the isotherm of phase transition.

\item  we restricted the expression for the chemical potential of the reference system
to $\beta\mu_0 = \xi$.
\end{itemize}

We postulated that the dependence ${\cal{E}}(\rho_{\rm max})={\cal{E}}(\Delta), $ being accurate
in the region of $Q>0$, remains also accurate for $Q<0$, if
the sum $\Delta = \Delta_\textrm{g}+\Delta_\textrm{l}$ is assumed as $\Delta$.

All the above demands a more detailed consideration.
The mutually-coordinated short-range and  long-range
interaction being taken into account will be one of the main tasks in further research.

\section*{Acknowledgements}

The author expresses his gratitude to I.M.~Mryglod and O.L.~Ivankiv for help, to
M.P.~Kozlovskii and all the researchers of the Department of the Statistical Theory of
Condensed Matter for useful discussions; to O.V.~Patsahan for the data on
the property of the parameter $\Delta$ in the critical region; to A.D.~Trokhimchuk for
discussions concerning the short-range order in condensed systems; to
Yu.V.~Holovach for proof-reading the paper and consistent remarks.
Our special thanks go to R.~Romanik for valuable considerations and the graphs,
and to V.O.~Kolomiets for collaboration.

\clearpage




%
%

\ukrainianpart
\title{Фазовий перехід 1-го роду в області критичної точки газ--рідина}
  \author{І.Р. Юхновський}
  \address{Інститут фізики конденсованих систем НАН України, вул. І.~Свєнціцького, 1,  79011 Львів, Україна}
%
\makeukrtitle
\begin{abstract}
Розглядається поведінка системи взаємодіючих частинок в області температур нижче критичної точки $T \leqslant T_\textrm{c}$. Завершується розрахунок великої статистичної суми, початий у попередніх роботах у методі колективних змінних. За базову густину міри береться четвірний (а не Гаусовий) розподіл. Описані події пов'язані з фазовим переходом 1-го роду, що відбуваються в результаті ізотермічного квазістатичного стиснення газу. Виділена лінія $\mu^*(\eta) =0$, на якій у газовій фазі під дією тиску виникає крапля рідини. Має місце рівність хімічних потенціалів газової і рідкої (у краплі) фаз; знайдено величину поверхневої енергії краплі, розраховано скриту роботу конденсації, визначено скачок густини, написано рівняння стану. Робота становить певну кінцеву стадію досліджень в області температур і густин, що включає у собі критичну точку рідина--газ.
\keywords критична точка, фазовий перехід першого роду, велика статистична сума, колективні змінні
\end{abstract}

\end{document}